\pdfoutput=1
\documentclass[11pt]{article}

\usepackage[preprint]{acl}

\usepackage{times}
\usepackage[T1]{fontenc}
\usepackage[utf8]{inputenc}

\usepackage{microtype}

\usepackage{inconsolata}

\usepackage{graphicx}

\usepackage{hyperref}
\usepackage{url}
\usepackage{soul}
\newcommand{\xmark}{\ding{55}}%
\usepackage{pifont}
\usepackage{xcolor}
\usepackage{colortbl}
\usepackage[most]{tcolorbox}
\usepackage{pifont}
\usepackage{algpseudocode}
\usepackage{amsfonts}
\usepackage{fdsymbol}
\let\oldcheckmark\checkmark
\renewcommand{\checkmark}{\ensuremath{\oldcheckmark}}
\usepackage{graphicx}
\usepackage{subfig}
\usepackage{mathtools}
\usepackage{amsmath}
\usepackage{amsthm}
\usepackage{listings}
\usepackage{wrapfig}
\definecolor{light_gray}{rgb}{.95,.95,.95}
\definecolor{custompurple}{RGB}{93,0,93}
\definecolor{customorange}{RGB}{255,132,6}
\definecolor{customgold}{RGB}{213,177,52}
\definecolor{customblue2}{RGB}{28,205,188}
\definecolor{no_persona_color}{RGB}{152,226,245}
\definecolor{persona_color}{RGB}{193,167,246}
\usepackage{booktabs}
\usepackage{adjustbox}
\usepackage{scalerel}
\usepackage{fontawesome5}
\usepackage{booktabs,tabularx,array,colortbl,xcolor}
\usepackage{siunitx}
\usepackage[most]{tcolorbox}
\usepackage{url}
\usepackage{twemojis}
\usepackage{dsfont}
\usepackage{bm}
\usepackage{bbm}
\usepackage{graphicx}
\usepackage{color}
\usepackage{multicol}
\usepackage{multirow}
\usepackage{wrapfig,lipsum,booktabs}
\usepackage[ruled,vlined,linesnumbered]{algorithm2e}
\usepackage{pgfplots}
\pgfplotsset{compat=1.12}
\usepackage{xcolor}
\usepackage{tikz}
\usepackage{xspace}
\usepackage{makecell}
\usepackage{soul}
\usepackage{amsmath,amsfonts}
\usepackage{subcaption}
\usetikzlibrary{calc}
\usepgfplotslibrary{groupplots}
\usetikzlibrary{angles,quotes} 
\usetikzlibrary{shapes,arrows}
\usetikzlibrary{backgrounds}
\usetikzlibrary{matrix}
\usepackage{tikz-3dplot}
\usepackage{hyperref}
\usepackage{cleveref}
\usepackage{paralist}
\usepackage{cancel}
\usepackage{xspace}
\usepackage{todonotes}
\usepackage{tabu}
\usepackage{rotating}
\usepackage{etoolbox}
\usepackage{adjustbox}
\usepackage{enumerate}
\usepackage{enumitem}
\setitemize{noitemsep,topsep=0pt,parsep=0pt,partopsep=0pt}
\setenumerate{noitemsep,topsep=0pt,parsep=0pt,partopsep=0pt}
\usepackage{pifont}
\usepackage{cancel}
\usepackage{lipsum}
\usepackage{listings,lstautogobble}
\usepackage{fancyvrb}
\usepackage{fvextra}
\usepackage{caption}
\usepackage{pgf-pie} 
\usepackage{array, makecell}
\usepackage{wrapfig}

\definecolor{codegreen}{rgb}{0,0.6,0}
\definecolor{codegray}{rgb}{0.5,0.5,0.5}
\definecolor{codepurple}{rgb}{0.58,0,0.82}
\definecolor{backcolour}{rgb}{0.95,0.95,0.92}

\definecolor{citysimlight}{RGB}{238,245,255}   % Light, airy blue background
\definecolor{citysimaccent}{RGB}{88,140,255}
\definecolor{CrowdCol}{HTML}{5A6EFF}   % vivid blue
\definecolor{ExpertCol}{HTML}{50F05A}  % bright green
\definecolor{GPTCol}{HTML}{FF2828}     % strong red
\definecolor{ClaudeCol}{HTML}{FA8C32}  % pumpkin orange

\definecolor{chart Idle}{gray}{.6}
\definecolor{chart Poor}{RGB}{242,28,28}
\definecolor{chart Ok}{RGB}{248,172,37}
\definecolor{chart Ideal}{RGB}{1,151,0}
\definecolor{chart Over}{RGB}{0,125,234}
\definecolor{customblues3}{HTML}{7EA8F0}

\newdimen\tempdim

\newcommand*{\ChartLegend}[1]{%
  \ifdim\lastkern=1sp %
    \hspace{1em}%
  \fi
  \ChartBox{0.75em}{#1}%
  \,#1%
  \kern-1sp\kern1sp\ignorespaces
}
\newcommand*{\ChartBox}[3]{%
  \begingroup
    \settoheight{\tempdim}{L}%
    \edef\tempheight{\the\tempdim}%
    \settodepth{\tempdim}{g}%
    \edef\tempdepth{\the\tempdim}%
    \tikz[
      baseline=0pt,
      inner sep=0pt,
    ]
    \node[
      fill={#3!#2},
      rounded corners=1pt,
      anchor=base,
    ]{%
      \vphantom{g\"A}%
      \pgfmathsetlength{\tempdim}{#1}%
      \kern\tempdim\relax
    };%
  \endgroup
}

\title{CityReal: Human-Aligned Urban Behavior and City Dynamics Simulation with Large-Scale LLM Agents}

\author{
 \textbf{Nicolas Bougie\textsuperscript{1}},
 \textbf{Xiaotong Ye\textsuperscript{1}},
 \textbf{Narimasa Watanabe\textsuperscript{1}}
 \\ \texttt{\{nicolas.bougie,tony.yip,narimasa.watanabe\}@woven.toyota}\\
\\
 \textsuperscript{1}Woven by Toyota
}

\begin{document}
\renewcommand*{\arraystretch}{1.2}
\newcommand*{\chart}[3]{%
  \ChartBox{20mm/3000*(#1-400)}{#2}{#3}%
}
\newcommand*{\humanchart}[3]{%
  \ChartBox{20mm/1300*(#1-700)}{#2}{#3}%
}
\maketitle
\begin{abstract}
Large-scale urban simulation plays a pivotal role in social science, traffic safety, and transportation policy. Recent work has shown that large language models, when prompted as agents, can generate lifelike daily routines at city scale. Yet these methods typically rely on few-shot prompting, causing agents to reproduce the LLM’s behavioral priors rather than the target population. We introduce CityReal, a modular framework for human-aligned urban simulation. \textsc{CityReal} models agents as intention-driven decision makers that pursue coherent mobility and activity plans rather than isolated step-by-step choices. They adapt over time by learning habits and preferences based on experience and constraints. To improve population-level realism, we learn textual adapters for behavior modules that align agent decisions with observed population statistics. Experiments show that \textsc{CityReal} improves alignment with real-world human behavior at both micro and macro levels. Scaling to tens of thousands of agents, it supports analysis of crowd density, place popularity, mobility flows, and well-being under different urban scenarios, offering a scalable testbed for urban simulation and forecasting.
\end{abstract}

\section{Introduction}
Reproducing the daily behavior of urban populations is a long-standing challenge in computational social science \citep{lazer2009computational,hofman2021integrating}, with applications ranging from traffic safety, transportation analysis, to urban planning. Traditional simulators typically model individuals through hand-crafted behavioral rules \citep{epstein1999agent,macal2005tutorial}. While transparent and scalable, these rules can make behavior rigid and difficult to adapt to unseen situations \citep{zheng2022ai,wang2023humanoid,feng2024citybench}. Thus, capturing the psychological, social, and environmental factors that shape urban behavior remains challenging. 

The emergence of large language models (LLMs) offers a new direction. Rather than relying on fixed decision rules, LLM-based agents can reason in natural language and choose actions conditioned on their demographic profile, internal state, and surrounding environment \citep{park2023generative,gao2024large}. Recent work has applied this paradigm to urban simulation. LLMob \citep{wang2024largelanguagemodelsurban} generates resident trajectories. AgentSociety \citep{piao2025agentsociety} studies collective social phenomena such as opinion polarization and policy response. CitySim \citep{bougie2025citysim} and MobileCity \citep{ye2025mobilecity} scale LLM-agent simulation to large populations with richer personas, memory, and internal needs.

Despite this progress, existing LLM-based urban simulators face several limitations. Prior approaches often rely on few-shot prompting to imitate plausible residents~\citep{wang2024largelanguagemodelsurban,bougie2025citysim}, without explicitly aligning agent decisions with observed human behavior. Besides, agents often make discretionary decisions as isolated choices, producing trajectories with limited continuity, whereas real behavior is organized around persistent intentions such as running errands or spending time in a particular area. Finally, agents rarely \emph{learn} from their own history. Their behavior is typically determined by the initial persona and prompts, limiting the emergence of habits, preferences, social tendencies, and practical constraints accumulated through experience. As a result, agents with similar demographic profiles may behave too similarly instead of developing distinct routines shaped by their own trajectories.

We propose CityReal, an LLM-based framework for simulating urban populations whose behavior is both coherent at the individual level and aligned with human patterns. First, we learn textual adapters with Monte Carlo Tree Search to calibrate agent decisions toward target population statistics while keeping the LLM frozen. Second, agents organize behavior around persistent intentions, maintaining stable motives across consecutive actions rather than selecting each activity independently. Third, CityReal introduces experience-driven adaptation through end-of-day reflection, allowing agents to gradually accumulate habits, preferences, social tendencies, and constraints from their own experience. Together with persona, memory, belief, mobility, and social modules, these mechanisms produce urban agents that can adapt to context while remaining calibrated to observed human behavior. This marks a paradigm shift in urban population simulation, enabling large-scale, human-aligned agent populations that capture the behavioral complexity and diversity of real-world societies.

\section{Related Work}

Replicating human behavior in urban environments remains a challenge ~\cite{hofman2021integrating,lazer2009computational}. Traditional agent-based models have been widely used to study complex phenomena, resource allocation, and policy evaluation~\cite{epstein1999agent,macal2005tutorial, wilensky2015introduction}. Yet, they depend on hard-coded rules or fixed utility functions, constraining their capacity to capture behavioral diversity, adaptability, and long-term dynamics~\cite{feng2024citybench,zheng2022ai,emnlp/WangCC23}. Frameworks such as CityBench~\cite{feng2024citybench} and AI4SIM~\cite{zheng2022ai} have attempted to integrate richer, data-driven approaches, yet realistic cognitive and motivational modeling remains limited. Recently, LLMs have opened new possibilities for simulating human-like agents in virtual worlds~\cite{park2023generative,gao2024large,li2023camel,wei2022chain, bougie2026perceptuillmagentshumanaligned}. LLM-powered agents can reason, plan, and interact through natural language~\cite{gao2024large,wei2022chain,bougie2025simuser,park2023generative, bougie2024generative}. \citet{iclr/HongZCZCWZWYLZR24} demonstrates how agents can collaborate in complex software engineering tasks. Few studies have also explored agent alignment, including user simulation for recommender-system evaluation~\cite{bougie2026beyond} and alignment to human interaction trajectories~\cite{bougie2026alignuser}. However, these methods depend on human behavior data for each simulated agent, making them difficult to scale to urban populations. As research moves toward larger-scale simulations, computational efficiency becomes crucial. \citet{corr/abs-2411-10109} scale up simulations to 1,000 agents, but still inherits prohibitive costs. To improve realism, AgentSociety~\cite{piao2025agentsociety} and MobileCity \cite{ye2025mobilecity} use episodic memory and gravity-based place selection. Similarly, CitySim \cite{bougie2025citysim} advances LLM-based urban simulation through recursive activity planning, dynamic memory, and belief modules. Yet existing approaches still struggle to align simulated populations with observed human behavior while preserving coherent and diverse individual trajectories. They also lack mechanisms for maintaining long-term behavioral coherence or allowing agents to develop habits and preferences from their own experiences.

\begin{figure*}[tbp]
    \centering
    \includegraphics[width=1.0\linewidth]{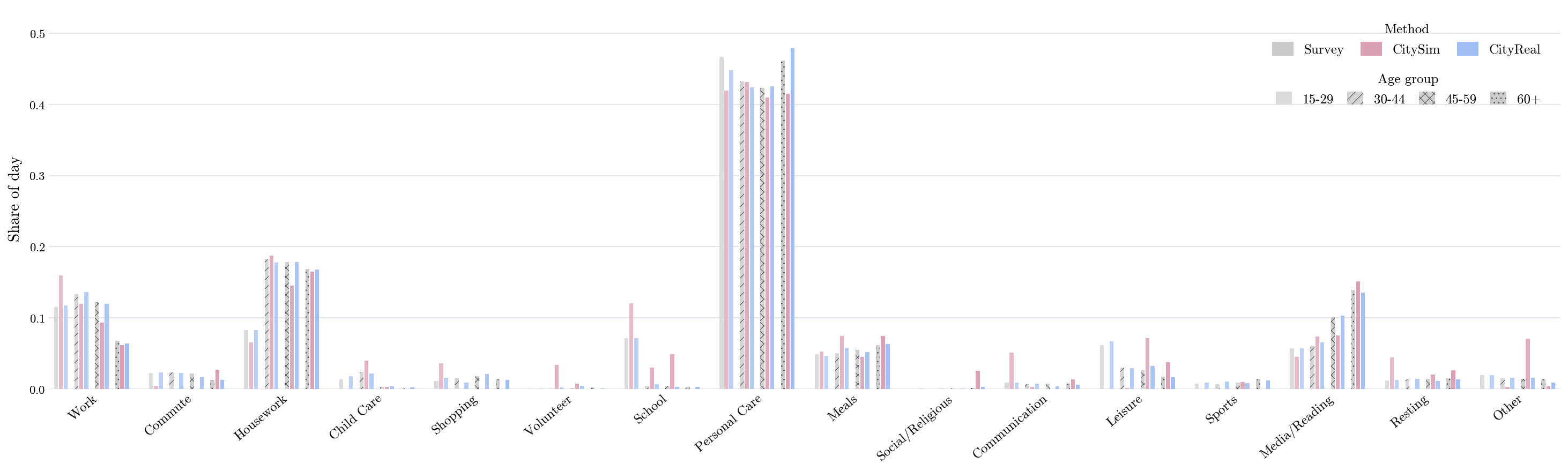}
    \caption{Time-use distribution across activity categories and age groups. Solid bars represent ground truth; striped bars show results from our simulation.}
    \label{fig:timeuse}
\end{figure*}

\section{Method}
\label{sec:method}
We propose \textbf{CityReal}, an LLM-based framework for simulating urban populations whose behavior is both individually coherent and aligned with observed real-world patterns. We model each resident as an agent embedded in a graph-structured city, where nodes correspond to places or areas and edges denote mobility connections (e.g., car, train). Each agent is initialized with a persona, after which CityReal calibrates behavior modules to target population statistics by learning textual adapters that condition their decisions. During simulation, an orchestrator selects the appropriate module from the agent's current context. Agents follow a perception--decision--reflection loop: they perceive the environment, retrieve relevant memories, continue or revise their behavior, and update habits, preferences, and beliefs through end-of-day reflection.

\subsection{Cognitive State Representation}
\label{sec:cognitive_state}

\noindent\textbf{Persona Module.} The persona encodes an agent's background: demographic attributes (age, gender, occupation, salary, household composition, life stage), spatial anchors (home, workplace, school), and psychographic traits (personality, lifestyle, routine tendencies). These attributes shape feasible activities and daily behavior (e.g., students follow school schedules, full-time workers commute on weekdays, and retirees maintain more flexible routines). 
% In particular, \texttt{salary} conditions economic behavior, such as spending decisions or destination choice.

\noindent\textbf{Memory Module.} The memory module connects past experience to future behavior through three components. The memory module contains three components. \textbf{Temporal memory} records experiences in chronological order, including time, location, activity, observation, and outcome. \textbf{Spatial memory} stores beliefs about places, such as affordability, convenience, crowding, atmosphere, and enjoyment; beliefs are updated after visits, while unvisited POIs are initialized from similar known places. \textbf{Reflective memory} stores conclusions distilled from experience. Unlike temporal memory, which logs what happened, reflective memory captures what the agent has learned and is retrieved during later decisions.

\noindent\textbf{Needs and Financial Module.} Agents track short-term needs (hunger, energy, safety, social connection, financial security) that evolve throughout the day and may interrupt ongoing routines. In this study, we explicitly model financial constraints through each agent's income, recent spending, and available budget, from which financial pressure is derived. This pressure directly shapes activity and destination choice, capturing socioeconomic heterogeneity.

\noindent\textbf{Belief Module.} Following CitySim \cite{bougie2025citysim}, the belief module maintains and updates the agent's subjective evaluations of places and social contacts. After each POI visit, the agent produces an assessment of the experience, covering  \texttt{affordability}, \texttt{convenience}, \texttt{crowding}, and \texttt{enjoyment}, which is integrated into spatial memory. For social contacts, beliefs over \texttt{affinity}, \texttt{trust}, and \texttt{familiarity} are revised after interactions, shaping future decisions such as whom to contact or whether to participate in group activities.

\subsection{Planning and Mobility}
\label{sec:planning_mobility}
Daily schedules are generated via a recursive decomposition of time into \texttt{[blocks]}, each including a starting time, duration, and activity or intention. At the beginning of the day, planning first assigns mandatory tasks (e.g., sleep, work), then recursively fills remaining \texttt{[EMPTY]} blocks with medium-priority tasks (e.g., meals, hygiene). 

\noindent The remaining flexible blocks are not pre-selected, instead, each is instantiated at execution time through the intention module. When a flexible block begins, the agent forms an intention --- a natural-language description of what it is trying to accomplish, including an optional area anchor, an expected duration, and a completion status. Examples include \textit{run errands near the station} or \textit{find an inexpensive meal before returning home}. The current intention provides a shared context for subsequent decisions, preventing flexible activities from being generated independently at each step. During execution, the orchestrator determines whether the active intention should be continued, revised, completed, or replaced.

\subsubsection{Place Selection}
\label{sec:place_selection}
When an activity requires movement, the agent selects a destination in two stages.

\noindent\textbf{Area Selection.} The agent first selects a broad area based on the current intention, location, schedule, beliefs, social context, and financial pressure. 

\noindent\textbf{POI Selection.} Within the selected area, candidate POIs are evaluated using a belief-aware gravity model:
\begin{equation}
w_{i,j}^{t} = \frac{\exp(\beta^\top f_{i,j}^{t})}{(1 + D_{i,j}^{t})^\gamma},
\end{equation}
where $D_{i,j}^{t}$ is the distance from agent $i$ to POI $j$, $f_{i,j}^{t} \in [0,1]^d$ is a vector of structured belief features (price, crowding, past satisfaction) derived from the agent's belief module, and $\gamma > 0$ controls distance sensitivity. The destination is sampled from the normalized distribution over candidates:

\begin{equation}
P(j \mid i, t) = \frac{w_{i,j}^{t}}{\sum_{k \in \mathcal{N}_i^t} w_{i,k}^{t}},
\end{equation}
where $\mathcal{N}_i^t$ denotes the candidate POI set.

\subsubsection{Transport Selection}
\label{sec:transport_selection}
After selecting a destination, the agent chooses a transport mode from available options: \{walking, bicycle, car, bus, or train\}, based on distance, time of day, weather, urgency, cost, accessibility, persona, and financial pressure. 

\subsection{Social Interaction}
\label{sec:social_module}
Following CitySim \cite{bougie2025citysim}, agents engage in both face-to-face and online interactions. Face-to-face interactions occur when agents are co-located, and the interaction is compatible with their intention, schedule, and needs, while online interactions occur during leisure or when social need is low. Partner selection is based on relationship strength and recent interaction history. Interaction outcomes update social beliefs (affinity, trust, and familiarity), which in turn shape future decisions such as whether to meet a contact.

\subsection{Experience-Driven Reflection}
\label{sec:reflection_module}
At the end of each day, this module extracts what the agent has learned from recent experiences and converts it into reusable behavioral insights. Rather than simply summarizing logs, the agent reflects on outcomes by considering what it liked or disliked, which constraints shaped its choices, and which habits, preferences, or tendencies emerged. These insights are retrieved and compared with the agent’s existing reflective memory, then integrated as gradual updates so that learned patterns reflect recurring evidence rather than isolated observations. Each reflection records the context in which a tendency applies, the behavioral tendency itself, supporting evidence, and its implications for future behavior. For example, an agent may learn to prefer quiet restaurants after work or leave early to avoid congestion.

\subsection{Population-Level Behavioral Alignment}                        \label{sec:alignment}  
LLM-based simulators are fundamentally limited by their reliance on prompting. Without explicit grounding, agents may reproduce the LLM's priors about human behavior rather than the population they are intended to simulate. We propose a calibration mechanism that steers agent decisions toward population-level distributions through textual adapters, without modifying the per-agent reasoning loop or retraining the underlying LLM.

\noindent\textbf{State.} For each agent $u$, we learn a natural-language adapter $\phi^{u,m}$ for each module $m \in \mathcal{M}$ that produce actions. Let $\Phi_t$ denote the collection of all adapters at iteration $t$. In the initial state $\Phi_0$, all adapters are empty and the simulator reduces to the vanilla agent prompts.                  

\noindent\textbf{Objective.} Let $\mathcal{K}$ index the set of population-level measures, and let $\delta_k(\Phi_t)$ denote the discrepancy between the simulated and target value of measure $k$ under adapter collection $\Phi_t$. We aggregate normalized discrepancies with a geometric mean, which encourages balanced alignment across measures and prevents a single metric from dominating the objective:
\begin{equation}
    R(\Phi_t) = \left(\prod_{k \in \mathcal{K}} 
    \frac{\delta_k(\Phi_t)}{\delta_k(\Phi_0) + \varepsilon}
    \right)^{1/|\mathcal{K}|} \cdot P_{\mathrm{plaus}} \cdot
    P_{\mathrm{div}}.
    \label{eq:r}
\end{equation}
\noindent To prevent trivial solutions, $R(\Phi_t)$ incorporates two penalties: $P_{\mathrm{plaus}}$ penalizes a collapse in individual agent plausibility, while $P_{\mathrm{div}}$ penalizes a loss of diversity within the population (see Appendix~\ref{app:reward}). Adapters are optimized iteratively, using the step return $r_t = R(\Phi_{t-1}) - R(\Phi_t)$ as the improvement signal at each iteration.

\noindent\textbf{Search.} Evaluating an adapter update requires running a simulation, hence exhaustive search is infeasible. We formulate adapter optimization as a tree search problem and use Monte Carlo Tree Search (MCTS) ~\cite{browne2012survey}. At each node expansion, an analyst LLM examines the current gap between simulated and target population statistics and proposes a set of possible actions $(m^{\star}, \mathcal{G}, \Delta, g)$. MCTS maintains value estimates for explored branches, so promising edits discovered in earlier branches naturally inform the search in later iterations (Appendix~\ref{app:search}).

\noindent\textbf{Action and Transition.} After MCTS selects an action $a_t = (m^{\star}, \mathcal{G}, \Delta, g)$, the tuple specifies a target module $m^{\star}$, a group of agents $\mathcal{G}$ defined by persona attributes, a behavioral edit $\Delta$, and a short rationale $g$. $\Delta$ is drawn from a vocabulary of behavioral adjustments (Appendix~\ref{app:style-axes}). For each agent in $\mathcal{G}$, a rewriting LLM updates the corresponding adapter using the agent's current adapter, persona, $\Delta$, and $g$. The rewriter is constrained to produce soft behavioral tendencies. Outputs containing clock times, exact distances, or absolutisms (\emph{always}, \emph{never}, \emph{must}) are rejected and regenerated, so adapters bias the agent's decisions without overriding its situational context and internal states. Adapters for agents outside $\mathcal{G}$ and for modules other than $m^{\star}$ are held fixed, so each action induces a sparse perturbation of $\Phi_t$.

\noindent\textbf{Transfer.} To avoid running search over the full population, adapters are optimized on a subset $\mathcal{U}_s$ chosen to cover the main persona types. The learned adapters are then transferred to the \textit{uncalibrated} agents $\mathcal{U} \setminus \mathcal{U}_s$ by nearest-neighbor matching in persona embedding space.

\section{Experiments}
\textbf{Settings.} All agents are powered by the GPT-5.4-mini version of ChatGPT, except when specified differently, with the number of agents set to 3,000 located in the Tokyo metropolitan area.
\\\textbf{Baselines.} We compare CityReal with GeAn \cite{uist/ParkOCMLB23}, AGA \cite{corr/abs-2402-02053}, HumanoidAgent \cite{emnlp/WangCC23}, and MobileCity \cite{ye2025mobilecity}. We also report results with our closest competitors, AgentSociety \cite{piao2025agentsociety}, and CitySim \cite{bougie2025citysim}.

\subsection{Macro-level Time Use}
\label{sec:exp_timeuse}

We assess whether CityReal reproduces realistic population-level activity patterns by comparing simulated time use against the 2021 Japanese national time use survey~\cite{e-stat2021timeuse}. Agents run for two simulated months, and their activities are mapped to survey categories (Work, Commute, Housework, Personal Care \& Sleep). We report the share of daily time spent in each category by age group. As illustrated in Figure~\ref{fig:timeuse}, CityReal closely matches the survey distribution, suggesting that the alignment stage calibrates towards the population-level statistics.

\subsection{Pairwise Human Preferences}
\label{sec:parwise_likeliness}
\begin{figure}[tbp]
    \centering
    \includegraphics[width=1.0\linewidth]{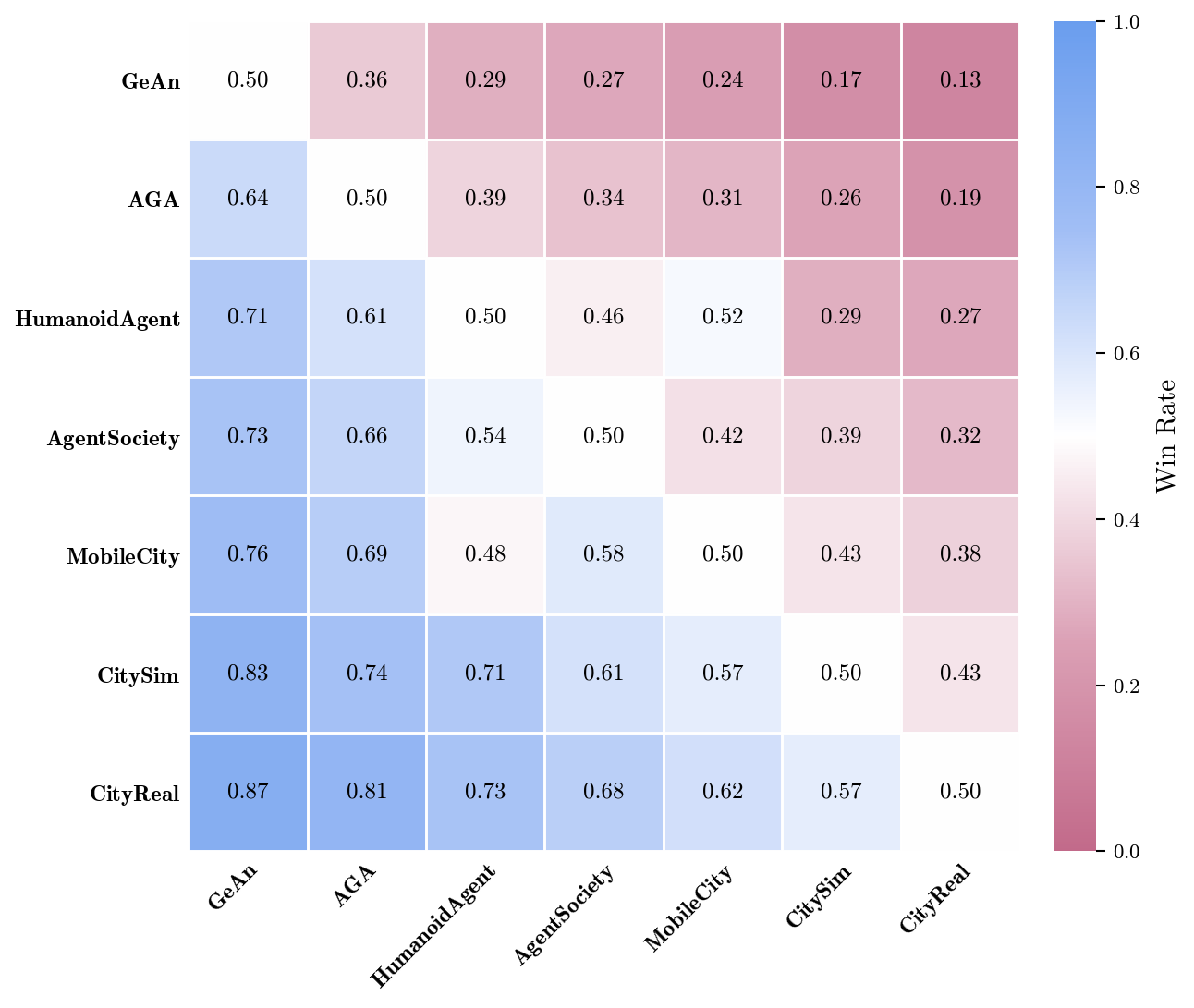}
    \caption{Pairwise win rate matrix. Each entry denotes the proportion of trials in which the row agent is judged more human-like than the column agent.}
    \label{fig:winrate}
\end{figure}

Beyond macro-level alignment, a realistic simulator should also produce human-like behavior at the individual level. Thus, we compare agents through pairwise judgments of anonymized daily routines, using 15 independent trials per approach. Outputs are first normalized with Llama-3.1 70B to mitigate stylistic bias, then judged by GPT-5 along three criteria: (i)~\textbf{Naturalness}, (ii)~\textbf{Coherence}, and (iii)~\textbf{Plausibility}. The pairwise win rate measures how often each agent is judged more human-like than another. Figure~\ref{fig:winrate} shows that CityReal achieves the highest average win rate across pairwise comparisons. This can be attributed to the reflective memory and explicit intention modeling, producing more socioeconomically consistent routines. MobileCity and AgentSociety produce more rigid and repetitive schedules, often overlooking social norms, raising suspicions of AI involvement.

\subsection{Travel Patterns}
\begin{figure}[tbp]
    \centering
    \includegraphics[width=1.0\linewidth]{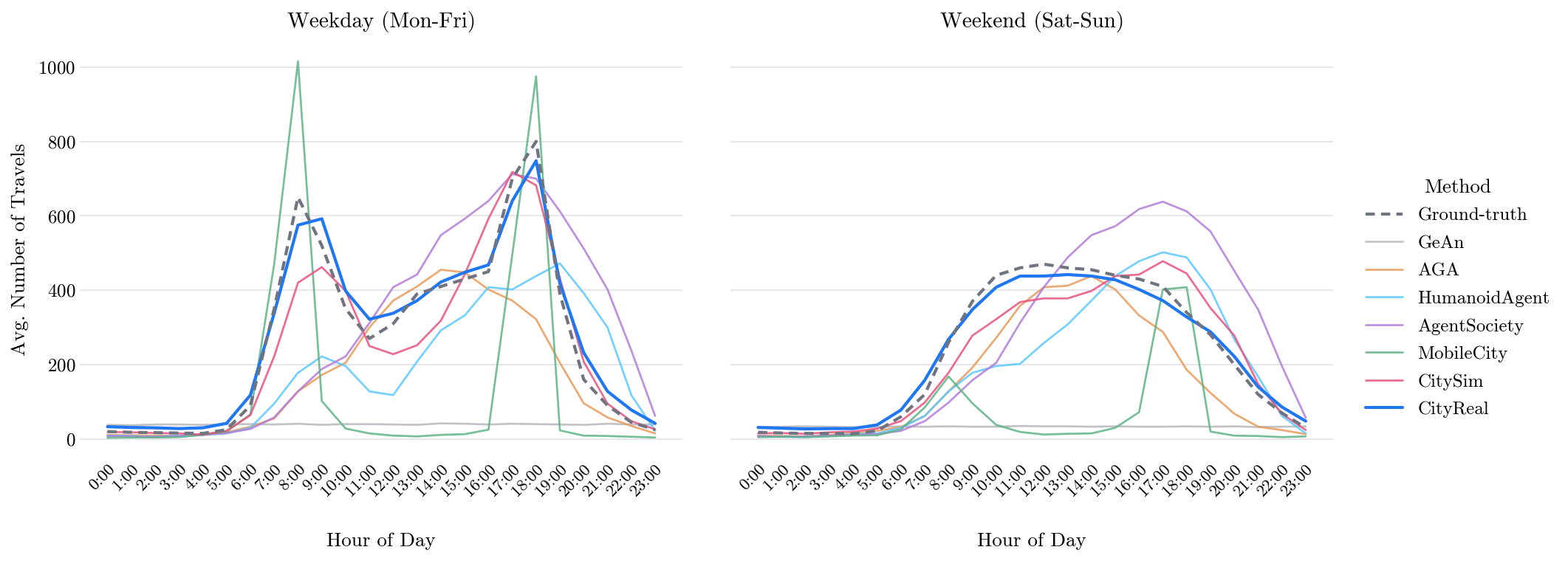}
    \caption{Average number of agent travels per hour on weekdays (left) and weekends (right).}
    \label{fig:mobility}
\end{figure}

Next, we compare simulated travel distributions against a proprietary city-scale dataset. Figure~\ref{fig:mobility} depicts the average number of trips per hour on weekdays and weekends. Prior methods generate travel decisions mainly from prompt-driven agent decisions. On the other hand, our alignment stage explicitly calibrates agent behavior against observed data. At the individual level, mandatory schedule planning produces consistent commute peaks. As a result, the proposed method closely reproduces both the timing and amplitude of commute peaks on weekdays and the gradual leisure travel patterns on weekends. In contrast, MobileCity produces unrealistically sharp peaks. Other LLM-based approaches capture broad temporal trends but fail to reproduce the precise timing and magnitude of travel peaks.

\subsection{Predicting POI Popularity}
\begin{figure}[tbp]
    \centering
    \includegraphics[width=1.0\linewidth]{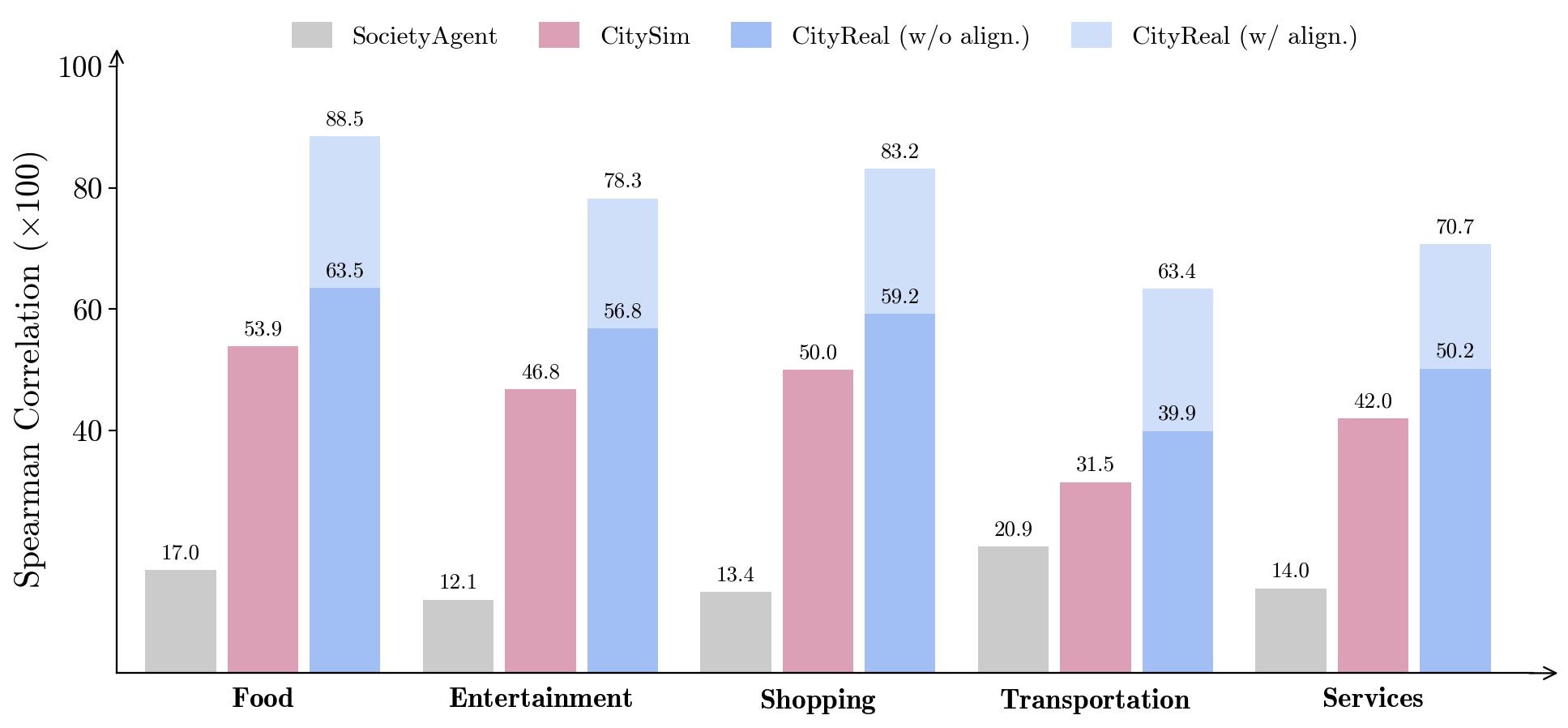}
    \caption{Comparison of real-world POI popularity and simulated visits in Shibuya.}
    \label{fig:poicorr}
\end{figure}
Predicting which POIs attract the most visitors is central to applications such as urban planning, retail strategy, and event management. We evaluate CityReal as a predictive tool for POI popularity in Shibuya, Tokyo. Ground truth is estimated from Google Maps ratings, while simulated popularity is measured by counting agent visits over one simulated month. We report Spearman rank correlations between simulated and real-world popularity. Figure~\ref{fig:poicorr} indicates that CityReal achieves stronger correlation with real-world popularity than CitySim. One reason is explicit financial constraints, which produce heterogeneous visits across income groups, preventing agents from disproportionately concentrating on highly rated or popular POIs.

\subsection{Social Studies using Synthetic Agents}
\begin{table}[tbp]
\centering
\small
\begin{tabular}{lcl}
\toprule
                 & F1-macro (mean $\pm$ std) &   \\
\midrule
GeAn           & 0.20 $\pm$ 0.03  & \chart{490}{32}{cyan} \\
AGA            & 0.21 $\pm$ 0.04  & \chart{518}{38}{magenta} \\
HumanoidAgent  & 0.24 $\pm$ 0.03  & \chart{600}{53}{yellow} \\
AgentSociety   & 0.29 $\pm$ 0.02  & \chart{718}{83}{magenta} \\
MobileCity     & 0.22 $\pm$ 0.03  & \chart{548}{45}{green} \\
CitySim   & 0.38 $\pm$ 0.02  & \chart{948}{100}{yellow}  \texttwemoji{3rd_place_medal}\\
\rowcolor{blue!10}
CityReal   & 0.43 $\pm$ 0.03  & \chart{1072}{100}{cyan} \texttwemoji{2nd_place_medal} \\
XGBoost & 0.45 $\pm$ 0.04 & \chart{1120}{100}{cyan} \texttwemoji{1st_place_medal} \\
\bottomrule
\end{tabular}
\caption{
Macro F1-score for well-being class prediction (5-class) across models, evaluated on a proprietary survey. Medals indicate top-3 methods.
}
\label{tab:wellbeing_ablation}
\end{table}

This experiment evaluates CityReal on population well-being estimation using a proprietary dataset of 1,200 survey responses collected in Japan, covering five well-being categories. Agents are initialized with persona profiles matching real respondents and simulate three weeks of daily activity. They then answer the same questionnaire, drawing on their accumulated memories. We compare against an XGBoost baseline trained on real activity and location data, as well as prior agent-based methods. Table~\ref{tab:wellbeing_ablation} highlights that our method outperforms all agent-based baselines, while XGBoost achieves the highest macro F1-score. These findings suggest that agent-based urban simulation can support population-level well-being estimation as a scalable and cost-efficient complement to conventional survey-based analysis.

\subsection{Modeling Crowd Density}
\begin{figure}[tbp]
\centering
\includegraphics[width=1.0\linewidth]{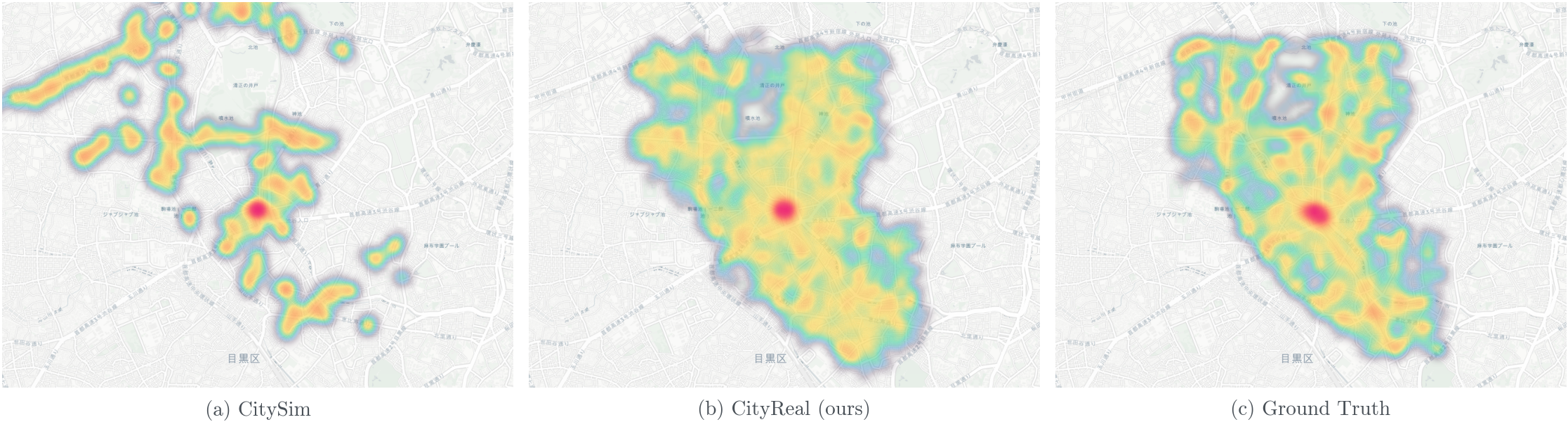}
\caption{Comparison of simulated (left) and real-world (right) crowd density heatmaps in Shibuya, Tokyo. Warmer colors indicate higher densities.}
\label{fig:density_heatmap}
\end{figure}

Predicting spatial crowd density is central to urban management, public safety, and event planning. We assess CityReal's ability to reproduce pedestrian concentration patterns across Shibuya, Tokyo. Agent visits are aggregated by location to generate simulated density heatmaps, which are compared against ground-truth distributions estimated from smartphone location data. As shown in Figure~\ref{fig:density_heatmap}, CityReal accurately reproduces real-world density patterns, with the highest concentrations around the train station and major commercial streets. This is partly driven by the intention module, which keeps agents within a target area and encourages sequential visits to nearby POIs, rather than independent destination choices after each activity. Experience-driven updates further diversify visitation patterns by making future choices depend on past outcomes rather than only on LLM priors.

\section{Conclusion}
We introduced CityReal, a modular framework for simulating human-aligned urban behavior with LLM-powered agents. Unlike prompt-driven agent simulators, CityReal targets both individual-level coherence and population-level alignment. It maintains intentions across consecutive decisions, allows agents to evolve from experience, and calibrates behavior-generating modules to match real-world statistics through learned adapters. Results demonstrate that our agents closely align with their human counterparts at both micro and macro levels. CityReal enables the study of complex urban phenomena and supports more realistic agent behaviors than prior agent-based models. These findings highlight our approach as a robust foundation for research and industry applications at the intersection of behavioral modeling, urban planning, and social studies.

\section{Limitations}
Although CityReal achieves strong performance across tasks, several limitations remain. First, the reproducibility of some experiments is constrained by the use of non-public datasets, which may limit direct comparison and independent verification. However, because most evaluations are based on macro-level aggregate statistics, future methods can still compare against CityReal using the reported targets. Second, our framework relies on large language models, hence the generated behaviors may reflect cultural, gender, socioeconomic, or other biases present in the underlying models and their training data. Relatedly, we observed occasional hallucinations when agents generated appraisals of recent, uncommon, or less popular POIs, which may introduce errors into downstream simulation outcomes. Third, the quality of the simulated behavior depends on the capabilities and failure modes of the LLM used as the agent backbone. Besides, the alignment process itself may introduce or reinforce biases if the target statistics, calibration data, or textual adapters encode incomplete or skewed assumptions about the population being simulated. Finally, CityReal consists of multiple interacting modules, making it challenging to fully isolate the contribution of each component. We provide ablation studies in the Appendix to partially address this issue, but a more fine-grained analysis of module interactions remains an important direction for future work.

\section{Ethics Statement}
This paper presents an LLM-driven framework for simulating urban human behavior at scale. Such simulations can support the study of city dynamics, mobility patterns, and social behavior in a scalable and cost-effective way. However, the use of synthetic urban agents also introduces important ethical considerations.

First, synthetic agents may reproduce or amplify biases related to age, gender, occupation, income, lifestyle, or other demographic factors if such biases are present in the underlying LLM, the agent initialization process, or the data used for calibration. If simulation outputs are used to inform urban planning or policy decisions, these biases could lead to analyses that overrepresent some groups while underrepresenting or disadvantaging others. Second, large-scale simulations of human behavior may be used to identify behavioral patterns that could support interventions aimed at steering collective behavior. Without appropriate transparency and oversight, such uses could raise concerns about consent, autonomy, and the potential manipulation of residents or communities.

Third, while synthetic agents are useful for early-stage exploration and low-cost evaluation of urban scenarios, they should not be treated as a substitute for real residents, stakeholders, or domain experts. Simulated behavior necessarily abstracts away many aspects of human experience, including lived experience, local knowledge, and contextual factors that may not be captured by an LLM-based system. We therefore recommend that synthetic agents be used to complement, rather than replace, participatory design, empirical studies, and expert review, especially when simulation results may influence real-world policies or interventions.

By making these limitations explicit, we aim to encourage the responsible use of LLM-based urban simulations. In particular, we believe such systems should be developed and deployed with attention to transparency, bias evaluation, human oversight, and the social consequences of decisions informed by synthetic populations.

\bibliography{custom}
\clearpage
\appendix

\begin{figure*}[tbp]
    \centering
    \includegraphics[width=0.90\linewidth]{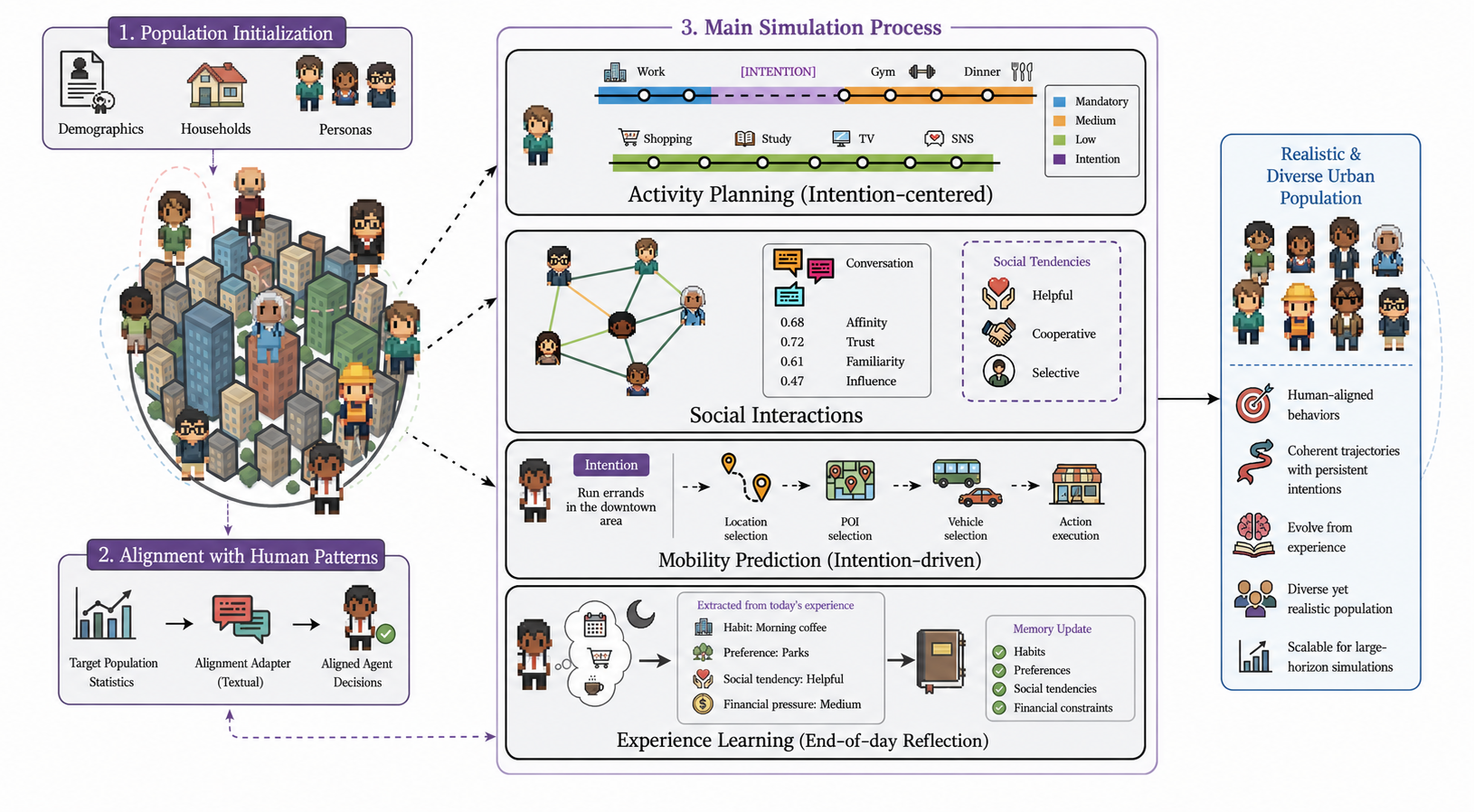}
    \caption{Overview of CityReal: LLM-based agents with diverse personas plan daily activities, interact socially, and navigate a virtual city environment.}
    \label{fig:overall_method2}
\end{figure*}

\section{Experimental Setup}
\label{app:experimental_setup}

\begin{figure}[tbp]
    \centering
    \includegraphics[width=1.0\linewidth]{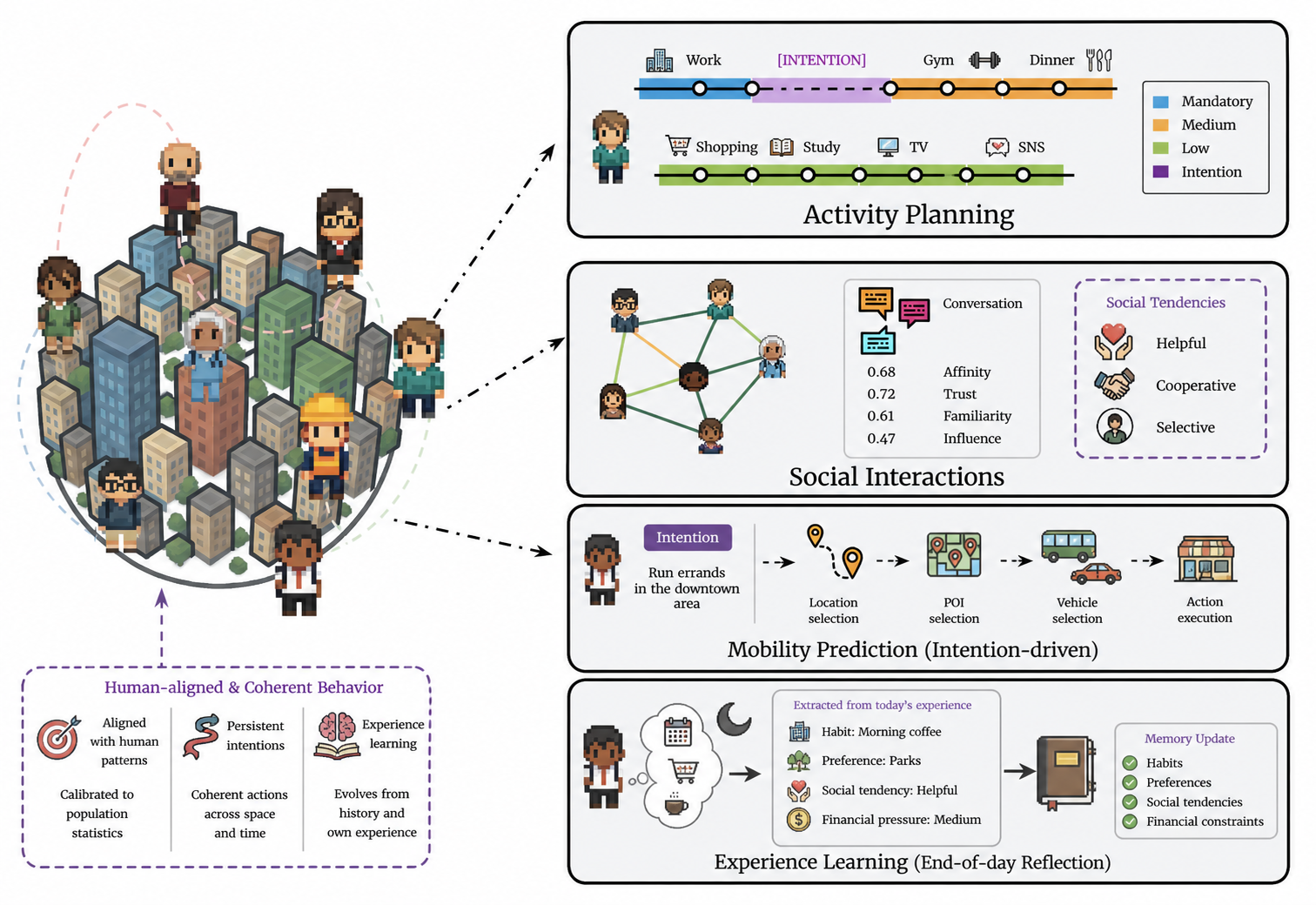}
    \caption{Overview of CityReal.}
    \label{fig:overall_method}
\end{figure}

Figure \ref{fig:overall_method2} presents an overview of CityReal. All agent attributes in the persona module are initialized from a proprietary survey-based dataset conducted in Japan. The attribute distributions closely match those observed in recent Japanese census statistics and lifestyle surveys~\cite{e-stat2021timeuse}. Each persona includes demographic attributes, household composition, life stage, occupation, salary, home and workplace or school locations, and psychographic traits. Big Five personality traits are discretized on a 3-point scale (1=low, 2=medium, 3=high). Home and workplace or school locations are assigned according to Japanese population density and OpenStreetMap data~\cite{openstreetmap}, ensuring realistic spatial distributions and feasible commutes. For reproducibility, researchers without access to our proprietary persona dataset may alternatively initialize agents from the open \textsc{Nemotron-Personas-Japan} dataset ~\citep{nvidia2025nemotronpersonasjapan}, which provides similar synthetic Japanese personas grounded in real-world demographic, geographic, and personality-trait distributions.

\textsc{CityReal} calibrates agent behavior using population-level textual adapters, see Figure \ref{fig:alignement}. Before running the full simulation, the alignment procedure searches for concise adapters attached to behavior modules such as daily planning, intention formation, place selection, transport selection, and reflection. These adapters are soft instructions that steer decisions toward target population statistics without overriding each agent's persona, memory, needs, location, social context, or financial state. Once learned, adapters are reused across agents, simulated days, and counterfactual scenarios. Note that the orchestrator module is not included in the search. 

The temporal memory retrieves the top $k_1=5$ entries from the past $\Delta t=24$ hours using cosine embedding similarity. Each memory entry stores \{time, location, activity, observation, outcome, key\}. Spatial memory stores beliefs about visited places, including affordability, convenience, crowding, and enjoyment. Beliefs for unvisited POIs are initialized from similar known places, using the $k=10$ most similar visited locations based on embedding distance. After each visit, the belief module produces a structured feedback of the experience and updates the corresponding spatial belief. Social beliefs over affinity, trust, and familiarity are updated after interactions.

Agents track short-term needs, including hunger, energy, safety, social connection, and financial security. Need thresholds for action interruption are $T_{\text{hunger}}=0.3$, $T_{\text{energy}}=0.3$, $T_{\text{safety}}=0.2$, and $T_{\text{social}}=0.2$, with priority order \textit{hunger} $>$ \textit{safety} $>$ \textit{energy} $>$ \textit{social}. Financial security is modeled separately from the immediate interruption needs. Rather than using a fixed interruption threshold, we represent it as a continuous financial pressure signal that conditions activity choice, destination choice, transport selection, and stay-home decisions. Let $B_u$ denote agent $u$'s monthly discretionary budget, $S_u^t$ its cumulative discretionary spending up to time $t$, and $A_u^t=\max(B_u-S_u^t,0)$ its remaining available budget. Let $\rho_t \in [0,1]$ denote the fraction of the month elapsed. We define financial pressure as:
\begin{equation}
p_u^t =
\mathrm{clip}\left(
\frac{S_u^t}{B_u+\epsilon} - \rho_t,\,
0,\,1
\right),
\end{equation}
where higher values indicate that the agent is spending faster than expected for the current point in the month. Similarly, pressure increases when the remaining budget $A_u^t$ is lower than the expected remaining budget $(1-\rho_t)B_u$. This signal biases agents toward lower-cost POIs, cheaper transport modes, fewer paid leisure activities, or staying at the current location, but it does not directly interrupt ongoing behavior.

The simulation operates with a 5-minute timestep, and all random seeds are fixed for reproducibility. Note that frontier LLM APIs may still exhibit minor nondeterminism due to backend-level implementation details. In detail, daily schedules are constructed from time blocks with a minimum granularity of 5 minutes, matching the resolution of human routine reporting in time-use surveys.

For pairwise human preference evaluation, we define the following criteria: (i) \texttt{Naturalness}: the extent to which actions align with the agent's profile, habits, constraints, and context; (ii) \texttt{Coherence}: the logical progression and goal-directedness of activities across time; and (iii) \texttt{Plausibility}: the overall believability of the sequence given realistic urban behavior.

\subsection{Module Details}
\label{app:module_details}

We provide additional implementation details for the main modules of \textsc{CityReal}.

\subsubsection{Planning and Intention Module}
\label{app:planning_intention_details}

Daily planning follows a recursive block-based procedure. Starting from an empty day, the planner first assigns fixed activities, such as sleep, work, school, medical appointments, or scheduled commitments, using the agent's persona, occupation, needs, and constraints. If a selected activity does not fill the entire interval, the block is subdivided according to the activity duration. The planner then recursively fills remaining \texttt{[EMPTY]} blocks with routine activities, such as meals, hygiene, and essential errands. Similar to human routines, flexible blocks are not fixed in advance. When a flexible block begins, the agent forms an intention: a compact natural-language description of what it aims to accomplish, optionally including an area anchor, expected duration, and completion status. Examples include \textit{run errands near the station} or \textit{find an inexpensive meal before returning home}. The active intention provides shared context for subsequent decisions and prevents flexible activities from being generated as disconnected one-step choices. 

During simulation, the orchestrator is invoked whenever a new decision is required, such as at the start or end of an activity, upon arrival at a destination, when a need becomes urgent, or when the environment changes. It receives the agent's current schedule block, active intention, location, retrieved memories, needs, financial pressure, nearby POIs, nearby agents, and weather. The orchestrator then decides whether to \texttt{continue}, \texttt{revise}, \texttt{complete}, or \texttt{replace} the current intention. The resulting action and observation are stored in temporal memory and used to update spatial beliefs, social beliefs, needs, and financial state.

For place selection, we use the two-stage area--POI procedure described in Section~\ref{sec:place_selection}. In all experiments, the area selector considers the top 10 candidate areas by relevance and proximity, and the POI selector considers up to 200 candidate POIs within the selected area. The belief-aware gravity model uses $\gamma=2.0$ for distance decay. The feature vector $f_{i,j}^{t}$ contains normalized estimates of affordability, crowding, satisfaction, and convenience from the agent's spatial memory. For unvisited POIs, these features are averaged over similar visited locations.

For transport selection, available modes encompass: walking, bicycle, car, bus, and train. The transport module conditions on distance, expected travel time, weather, urgency, cost, accessibility, persona constraints, and financial pressure. Face-to-face interactions are limited to one partner per 30-minute window to avoid unrealistically frequent social behavior. Online interactions may occur during leisure periods or when agents seek social contact without traveling.

\subsubsection{Reflection Module}
\label{app:reflection_details}
At the end of each simulated day, \textsc{CityReal} applies the reflection module once for each agent. The input consists of a compact summary of the day's trajectory, up to 24 representative temporal memories from that day, and the top-8 most relevant existing reflective memories retrieved by semantic similarity. The model generates up to four candidate reflections using a fixed schema: the context in which a tendency applies, the behavioral tendency itself, supporting evidence, and its implication for future behavior. Each candidate is compared with the retrieved reflective memories. If it is consistent with an existing reflection, the entries are consolidated by refining the applicable context and adding the new evidence. If it captures a distinct recurring pattern, it is stored as a separate reflection. At most three new reflective entries are added per day; candidates with limited evidence are not added as standalone entries. During simulation, downstream modules retrieve the top-5 relevant reflective memories. To keep memory bounded, each agent maintains at most 100 reflective entries, with older or redundant entries summarized when the limit is exceeded.

\section{Population-Level Alignment}
\label{app:alignment_details}
\begin{figure}[tbp]
    \centering
    \includegraphics[width=1.0\linewidth]{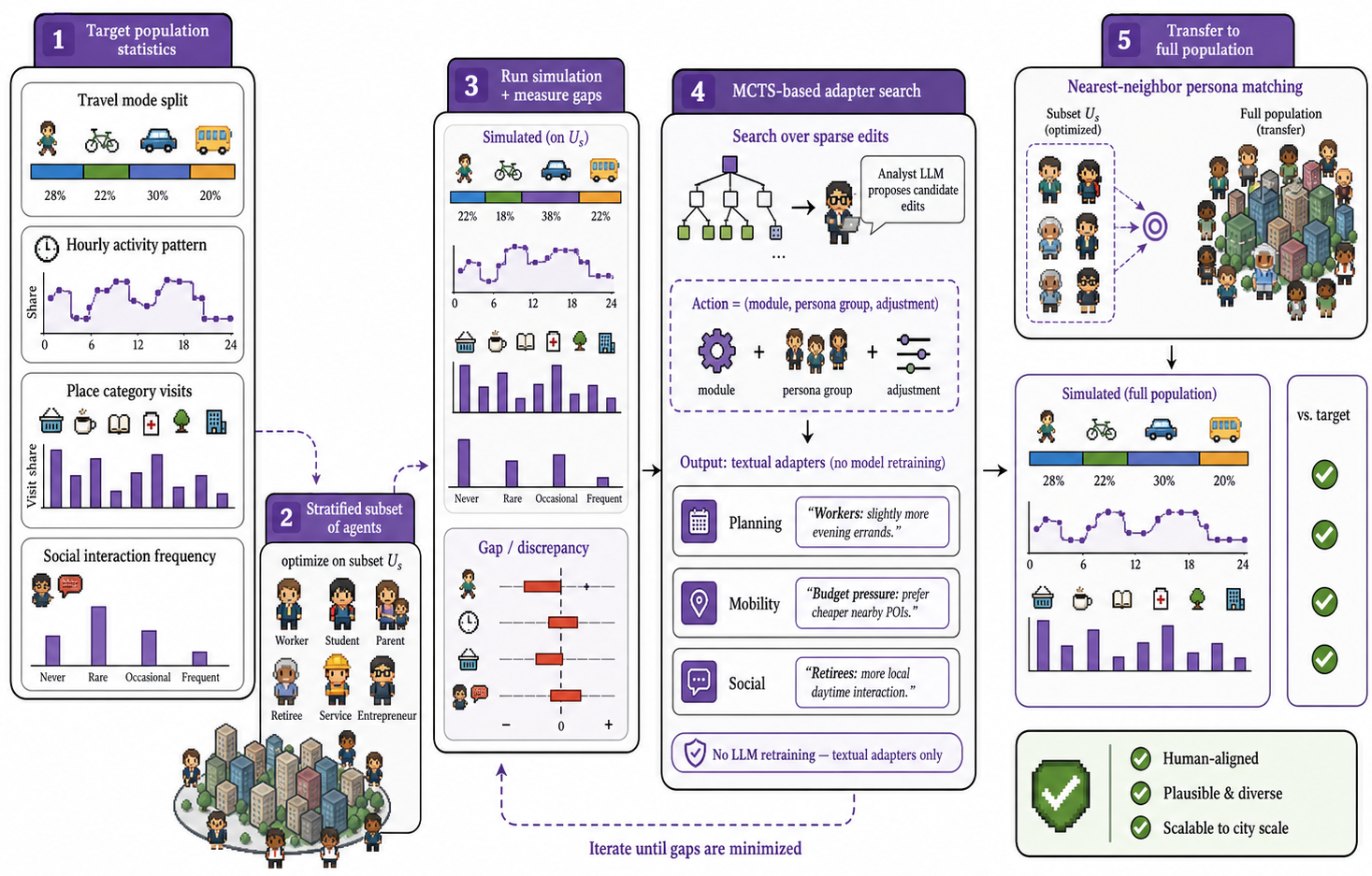}
    \caption{Population-level alignment via MCTS adapter search. An analyst LLM proposes edits to targeted persona groups; a rewriter proposes new textual adapters. A rollout on the subset of agent scores the new adapters, and MCTS uses this score to pick the next edit.}
    \label{fig:alignement}
\end{figure}

\textsc{CityReal} aligns agent behavior with population-level statistics through textual adapters (Figure \ref{fig:alignement}). For each agent $u$ and behavior module $m$, the adapter $\phi^{u,m}$ is a short natural-language instruction appended to the module prompt. Let $\Phi_t$ denote the collection of adapters at search iteration $t$. In the initial state $\Phi_0$, all adapters are empty and the simulator reduces to the unadapted version. The alignment objective compares simulated aggregate statistics with target population statistics. Let $\mathcal{K}$ denote the set of available measures and let $\delta_k(\Phi_t)$ be the discrepancy between the simulated and target value for measure $k$. We optimize the baseline-normalized objective:
\begin{equation}
    R(\Phi_t) = \left(\prod_{k \in \mathcal{K}}
    \frac{\delta_k(\Phi_t)}{\delta_k(\Phi_0) + \varepsilon}
    \right)^{1/|\mathcal{K}|}
    \cdot P_{\mathrm{plaus}}
    \cdot P_{\mathrm{div}},
\end{equation}
where $P_{\mathrm{plaus}}$ penalizes loss of individual plausibility and $P_{\mathrm{div}}$ penalizes collapse in population diversity. Lower values of $R(\Phi_t)$ indicate better alignment. The step return is defined as $r_t = R(\Phi_{t-1}) - R(\Phi_t)$, so positive return corresponds to a reduction in discrepancy.

\subsection{Alignment Objective Details}
\label{app:reward}

\paragraph{Per-Measure Discrepancy.}
The discrepancy term $\delta_k$ in Eq.~\eqref{eq:r} depends on the type of population-level measure. For continuous distributions, such as travel distance and inter-trip interval, we use a log-space $1$-Wasserstein distance combined with a weighted CCDF $L_1$ term. For categorical distributions, such as transport distribution, origin--destination flows, and trip purpose, we use Jensen--Shannon divergence. For scalar summaries with dispersion, such as wake-up time, we average the relative error of the median and interquartile range.

\paragraph{Plausibility Penalty.}
This penalty prevents adapters from improving aggregate alignment by making individual agents behave unrealistically. Agents in the simulator emit an \emph{emergency-break} event whenever a critical need (sleep, food, health) forces them to abort their current activity. Let $\rho_t$ be the mean number of these events per agent-day under adapter collection $\Phi_t$, and $\rho_0$ the corresponding value under the unadapted simulator $\Phi_0$. We set the threshold relative to the baseline, $\rho_{\mathrm{thr}} = \max(1.5,\, \rho_0 + 1.0)$, and apply the penalty: \begin{equation}
   P_{\mathrm{plaus}} \;=\; 1 + \lambda_p \cdot \max\!\left(0,\; \frac{\rho_t - \rho_{\mathrm{thr}}}{\rho_{\mathrm{thr}} + \varepsilon}\right) 
\end{equation}
with $\lambda_p = 0.5$. Since lower $R(\Phi_t)$ is better, this term raises the objective whenever alignment causes excessive interruptions.

\paragraph{Diversity Penalty.}
The diversity penalty prevents adapters from matching aggregate statistics by making agents behave too similarly. We partition the population into \emph{groups}. Groups are constructed from bucketed persona and behavior features, including \texttt{age\_bin}, \texttt{household}, \texttt{occupation}, \texttt{trips\_per\_day}, and \texttt{local\_trip\_rate}. To ensure reliable group-level estimates, each group must contain at least five agents. For each group and module, we compute the Shannon entropy of agent decisions under the adapted simulator $\Phi_t$ and divide it by the corresponding entropy under the unadapted simulator $\Phi_0$. We then take the geometric mean of these entropy ratios across groups and modules, denoted by $h_t$. When $h_t < 0.80$, we apply the penalty $ P_{\mathrm{div}} = 1 + \lambda_d \max(0, 0.80 - h_t)$, with $\lambda_d=1.0$. Since $R(\Phi_t)$ is minimized, this term increases the objective when adaptation reduces behavioral diversity by more than $20\%$ relative to the unadapted agents.

\subsection{Adapter Edit Vocabulary}
\label{app:style-axes}

Table~\ref{tab:axes} lists examples of the edit axes used by the analyst during adapter search. Each search action selects one target module and up to two axis--direction pairs. The rewriter then converts the selected edit into a textual adapter.  
  
\begin{table}[h]
\centering
\small 
\resizebox{1.0\linewidth}{!}{
\begin{tabular}{@{}lll@{}}
\toprule
\textbf{Module} & \textbf{Axis} & \textbf{Directions} \\
\midrule
\multirow{4}{*}{daily structure}
& regularity                  & lower / higher \\
& wake\_shift                 & earlier / later \\
& evening\_activity\_tendency & lower / higher \\
& weekday\_rigidity           & lower / higher \\
\midrule
\multirow{4}{*}{intention}
& outing\_propensity          & lower / higher \\
& locality\_preference        & lower / higher \\
& routine\_preference         & lower / higher \\
& social\_outing\_tendency    & lower / higher \\
\midrule
\multirow{4}{*}{trip}
& exploration\_tendency       & lower / higher \\
& multi\_stop\_tendency       & lower / higher \\
& familiar\_place\_preference & lower / higher \\
& destination\_range\_tendency & shorter / longer \\
\midrule
\multirow{4}{*}{vehicle}
& walking\_preference         & lower / higher \\
& transit\_preference         & lower / higher \\
& effort\_aversion            & lower / higher \\
& weather\_sensitivity        & lower / higher \\
\midrule
\multirow{4}{*}{home activity}
& rest\_tendency              & lower / higher \\
& chore\_tendency             & lower / higher \\
& hobby\_tendency             & lower / higher \\
& family\_engagement          & lower / higher \\
\bottomrule 
\end{tabular}}                                          
\caption{Behavioral edit axes used for adapter edits. Each axis includes a \emph{neutral} option, indicating that the corresponding behavior should remain unchanged}                  
\label{tab:axes}                                                          \end{table}

\subsection{Search Procedure}                                                \label{app:search}
\begin{figure}[tbp]
    \centering
    \includegraphics[width=1.0\linewidth]{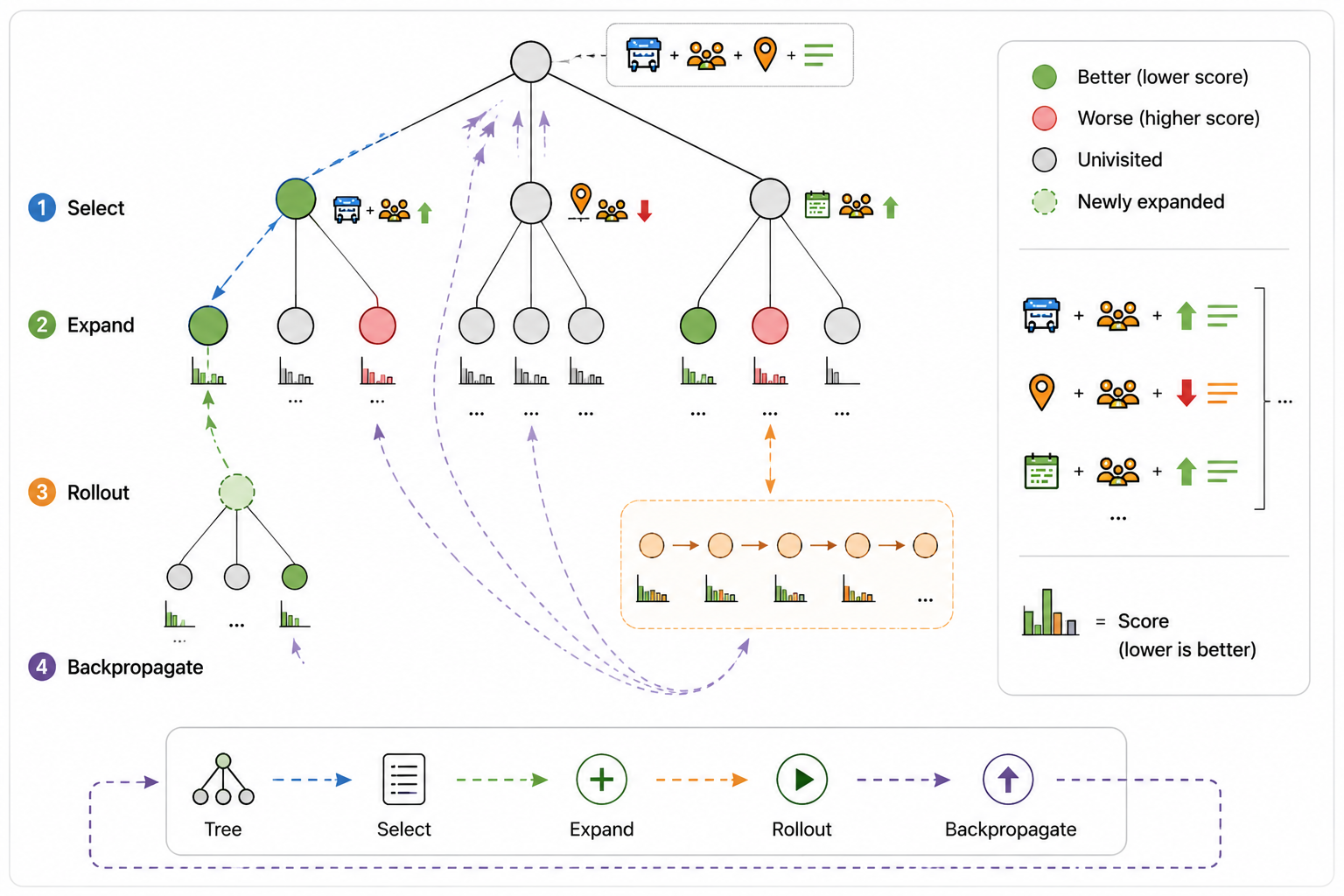}
    \caption{MCTS search procedure.}
    \label{fig:mcts_search}
\end{figure}

\paragraph{Algorithm.} At each iteration, we select a node from the root using UCT, expand it, evaluate each candidate through a rollout, and back-propagate cumulative returns along the evaluated trajectory (Figure~\ref{fig:mcts_search}). All evaluated candidates are added as sibling nodes, so candidates that are not selected immediately can still be revisited by UCT in later iterations. We maintain global statistics for each $(\text{module}, \text{edit})$ pair, allowing effective edits to generalize across agent groups.

\paragraph{Candidate Generation.} At each expansion step, an analyst LLM compares the simulated and target statistics and proposes candidate adapter edits. Each candidate specifies the module and agent group to adjust. We retain only candidates that use the allowed edit axes in Table~\ref{tab:axes} and cover a sufficiently large group.

\paragraph{Hyperparameters.} We utilize UCT constant $\sqrt{2}$, global exploration constant $1.0$, candidate pool size $3$, maximum tree depth $4$, and at most $20$ iterations. Search stops early when the best rollout reward fails to improve by more than $0.005$ over $5$ consecutive iterations. Analyst and rewriter LLMs use temperature $0.1$, rollout simulations employ temperature $0.2$.                                                                                                            
\paragraph{Subset Search.} To keep alignment efficient, we optimize adapters on a subset of $500$ agents covering major age, household, occupation, and spatial-anchor types. After search, adapters are transferred to the full population by matching each remaining agent to the closest subset agent in persona-embedding space.

\section{Discussion}
We acknowledge that our method exhibits certain limitations. CityReal agents produce collective behaviors consistent with established urban mobility patterns, yet the underlying reasoning of individual decisions remains partially opaque due to the black-box nature of large language models. A plausible explanation is that LLMs internalize behavioral patterns from training corpora covering urban routines and daily life across diverse global contexts. Disentangling learned cultural norms from genuine generalization remains an open challenge for LLM-based simulation.

\textsc{CityReal} models socioeconomic constraints through salary-conditioned personas, tracks spending, and estimates a financial-pressure signal. This improves behavioral grounding, but also makes the simulation dependent on the quality of the underlying economic data. When income or expenditure information is coarse, financial pressure may only approximate the constraints faced by real residents, which can bias conclusions about economically driven differences in behavior. 

Like other LLM-based simulators, \textsc{CityReal} may inherit demographic, cultural, or behavioral biases from the underlying model. Population-level alignment reduces this risk by grounding simulated behavior in observed aggregate statistics, and the financial-state module adds an explicit source of socioeconomic variation. However, aggregate statistics may still hide differences among minority or marginalized groups, especially when these groups are weakly represented in the calibration data. Although our experiments span multiple age groups, occupations, and income levels, evaluating alignment for underrepresented populations remains an important direction for future work.

CityReal uses a hybrid design in which LLMs handle decisions that require contextual reasoning, while explicit models track latent states such as financial pressure, beliefs, and short-term needs. This design improves both efficiency and interpretability. In preliminary experiments, LLMs were costly and unreliable for estimating scalar internal states, whereas explicit models were easier to inspect and control. Needs are modeled following established psychological frameworks~\cite{mcleod2007maslow}, and financial pressure is computed from tracked income and expenditure, making agent behavior more transparent and auditable.

Some experiments leverage GPT as an evaluator while the agents are generated with models from the same family, which may introduce stylistic bias in pairwise judgments. We mitigate this risk through cross-model validation with Gemini-3.5, but LLM-as-judge evaluation remains imperfect. Future work should incorporate larger-scale human evaluation and additional independent evaluators.

The alignment stage is essential for matching population-level statistics, but it depends on the quality of the calibration targets. If the target statistics are incomplete, biased, or too coarsely aggregated, the learned adapters may reproduce these limitations rather than recover the true behavioral distribution. Optimizing aggregate fit can also downweight rare but realistic behaviors, such as atypical routines or uncommon activity sequences. Although the plausibility and diversity penalties are designed to reduce this risk, preserving long-tail behavior remains challenging. Finally, adapters are tied to the target population and city context, and may need to be recalibrated when transferred to new urban environments.

\textsc{CityReal} also depends on sufficient local data for initialization. When demographic, spatial, or economic records are sparse, personas, spatial anchors, and financial states become less reliable. This weakens downstream modules such as belief formation and reflection, which rely on meaningful prior context and accumulated experience. Salary-informed initialization and empirical demographic distributions improve grounding, but they remain approximations of real individual profiles, especially for groups with limited observational coverage.

\section{Pseudo-Code}
\label{sec:pseudocode}
We summarize the main simulation loop of \textsc{CityReal} below:

\begin{tcolorbox}[
    colback=citysimlight,
    colframe=customblues3,
    boxrule=0.9pt,
    sharp corners=south,
    breakable=true,
    title= Algorithm 1: \textsc{CityReal} Simulation Loop,
    left=2mm, right=2mm, top=1mm, bottom=1mm
]
\label{alg:cityreal_loop}
\small

\textbf{Input:} city environment, agents, target population statistics\\
\textbf{Output:} trajectories and aggregate urban outcomes

\vspace{0.3em}
\texttt{calibrate\_modules()} \hspace{0.5em} \textit{// population-level textual adapters}

\vspace{0.1em}
\texttt{initialize\_agents()} \hspace{0.5em} \textit{// personas, needs, budgets, beliefs, memories}

\vspace{0.3em}
\textbf{For each day:}
\vspace{0.1em}

\hspace*{1.2em}\textbf{For each agent:}
\vspace{0.1em}

\hspace*{2.2em}\texttt{plan\_day()} \hspace{0.5em} \textit{// mandatory, routine, flexible blocks}

\vspace{0.2em}
\hspace*{1.2em}\textbf{For each time step:}
\vspace{0.1em}

\hspace*{2.2em}\textbf{For each active agent:}
\vspace{0.1em}

\hspace*{3.2em}\texttt{perceive(); retrieve\_memory()}
\vspace{0.1em}

\hspace*{3.2em}\texttt{update\_needs\_and\_financial\_pressure()}
\vspace{0.1em}

\hspace*{3.2em}decision $\leftarrow$ \texttt{orchestrator()}
\vspace{0.1em}

\hspace*{3.2em}\textit{// continue, revise, complete, or replace intention}

\vspace{0.2em}
\hspace*{3.2em}\textbf{if} decision.requires\_new\_activity:
\vspace{0.1em}

\hspace*{4.2em}intention $\leftarrow$ \texttt{update\_intention()}
\vspace{0.1em}

\hspace*{4.2em}activity $\leftarrow$ \texttt{select\_activity(intention)}
\vspace{0.1em}

\hspace*{4.2em}\textbf{if} activity.requires\_move:
\vspace{0.1em}

\hspace*{5.2em}area, poi $\leftarrow$ \texttt{select\_place(intention)}
\vspace{0.1em}

\hspace*{5.2em}transport $\leftarrow$ \texttt{select\_transport(poi)}
\vspace{0.1em}

\hspace*{5.2em}\texttt{move(poi, transport)}

\vspace{0.2em}
\hspace*{3.2em}\textbf{else:}
\vspace{0.1em}

\hspace*{4.2em}\texttt{continue\_current\_behavior()}

\vspace{0.2em}
\hspace*{3.2em}\texttt{social\_interaction()}
\vspace{0.1em}

\hspace*{3.2em}\texttt{execute\_activity()}
\vspace{0.1em}

\hspace*{3.2em}\texttt{update\_state()} \hspace{0.5em} \textit{// memory, beliefs, needs, spending}

\vspace{0.3em}
\hspace*{1.2em}\textbf{For each agent:}
\vspace{0.1em}

\hspace*{2.2em}\texttt{reflect\_end\_of\_day()} \hspace{0.5em} \textit{// habits, preferences, constraints, beliefs}

\vspace{0.2em}
\hspace*{1.2em}\texttt{aggregate\_outcomes()} \hspace{0.5em} \textit{// mobility, crowds, popularity, well-being}

\end{tcolorbox}

\section{Additional Experiments}

\subsection{Human Likeliness}
\begin{table}[btp]
\centering
\resizebox{1.0\columnwidth}{!}{
\begin{tabular}{lcccc}
\toprule
\textbf{Method} & \textbf{Activity} & \textbf{Dialogue} & \textbf{Mobility} & \textbf{Event Reaction} \\
\midrule
GeAn & 3.16 $\pm$ 0.19 & 3.97 $\pm$ 0.04 & 3.13 $\pm$ 0.16 & 3.06 $\pm$ 0.21 \\
AGA & 3.24 $\pm$ 0.27 & 3.95 $\pm$ 0.03 & 3.26 $\pm$ 0.24 & 3.17 $\pm$ 0.20 \\
HumanoidAgent & 3.38 $\pm$ 0.31 & 4.00 $\pm$ 0.06 & 3.30 $\pm$ 0.21 & 3.31 $\pm$ 0.17 \\
AgentSociety & 4.05 $\pm$ 0.23 & 4.09 $\pm$ 0.05 & 3.88 $\pm$ 0.25 & 3.86 $\pm$ 0.21 \\
MobileCity & 4.14 $\pm$ 0.26 & 4.05 $\pm$ 0.06 & 4.06 $\pm$ 0.19 & 3.83 $\pm$ 0.17 \\
CitySim & 4.40 $\pm$ 0.17 & 4.25 $\pm$ 0.04 & 4.23 $\pm$ 0.15 & 4.15 $\pm$ 0.15 \\
\rowcolor{blue!10}
CityReal w/o Align. & 4.47 $\pm$ 0.16 & 4.28 $\pm$ 0.05 & 4.33 $\pm$ 0.15 & 4.27 $\pm$ 0.14 \\
\rowcolor{blue!10}
CityReal & \textbf{4.56} $\pm$ \textbf{0.14} & \textbf{4.32} $\pm$ \textbf{0.05} & \textbf{4.46} $\pm$ \textbf{0.13} & \textbf{4.41} $\pm$ \textbf{0.13} \\
\bottomrule
\end{tabular}}
\caption{Human-likeness scores evaluated by GPT-5 across city simulation domains. Higher values indicate greater similarity to real human responses.}
\label{tab:llm_humanlike_city}
\end{table}

Following \citet{chiang2023can}, GPT-5 assesses whether agent behaviors appear human-generated or LLM-generated. For each method, we collect 20,000 outputs across four domains: daily activities, dialogue, mobility choices, and event reactions. Each sample is scored on a 5-point Likert scale, with higher scores indicating greater resemblance to human behavior. Table~\ref{tab:llm_humanlike_city} reports that \textsc{CityReal} achieves the highest scores across all domains. Gains are largest for mobility and event reaction, where persistent intentions and belief-aware mobility produce more coherent responses to changing contexts. This indicates that population-level adapters provide additional behavioral calibration, including at the agent level. Notably, the variant without alignment also outperforms \textsc{CitySim}. Dialogue gains are smaller, which is expected because dialogue fluency is already strong across recent LLM-based simulators. Overall, these results suggest that \textsc{CityReal} yields more human-like behavior by making agent decisions more coherent, context-sensitive, and faithful.

\subsection{Belief Estimation}

\begin{figure}[tbp]
    \centering
    \includegraphics[width=1.0\linewidth]{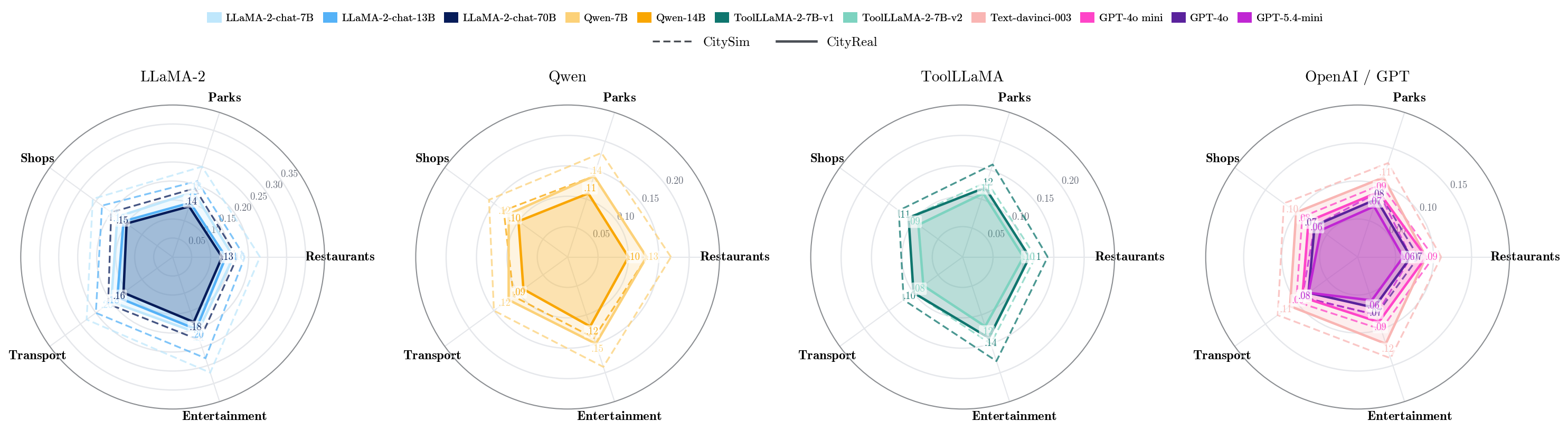}
    \caption{Category-wise mean absolute error (MAE) of belief estimation for unvisited POIs, evaluated across five semantic categories (\emph{Restaurants}, \emph{Parks}, \emph{Shops}, \emph{Transport}, \emph{Entertainment}) and eleven LLM models. Lower values indicate higher accuracy. }
    \label{fig:belief}
\end{figure}
Place beliefs directly condition destination choice and activity planning in CityReal. We evaluate belief estimation accuracy by initializing each agent with belief vectors derived from visited POIs, then tasking it with predicting beliefs for a disjoint set of unvisited POIs. For each test POI, we compute the mean absolute error (MAE) between predicted and ground-truth beliefs across five semantic categories. As shown in Figure~\ref{fig:belief}, larger models achieve lower errors overall, with GPT-5.4-mini performing best across categories, followed by GPT-4o mini and Qwen-14B. ToolLLaMA is competitive for \emph{Transport} and \emph{Shops}, while smaller models such as LLaMA-2 produce larger errors, especially for \emph{Entertainment}. This study indicates that stronger models provide more reliable place-belief estimates, which is important because \textsc{CityReal} uses these beliefs to condition destination choice, financial decisions, and long-term place preferences.

\subsection{Population-level Time-use Alignment}
\label{sec:exp_group_timeuse}

\begin{figure*}[tbp]
    \centering
    \includegraphics[width=1.0\linewidth]{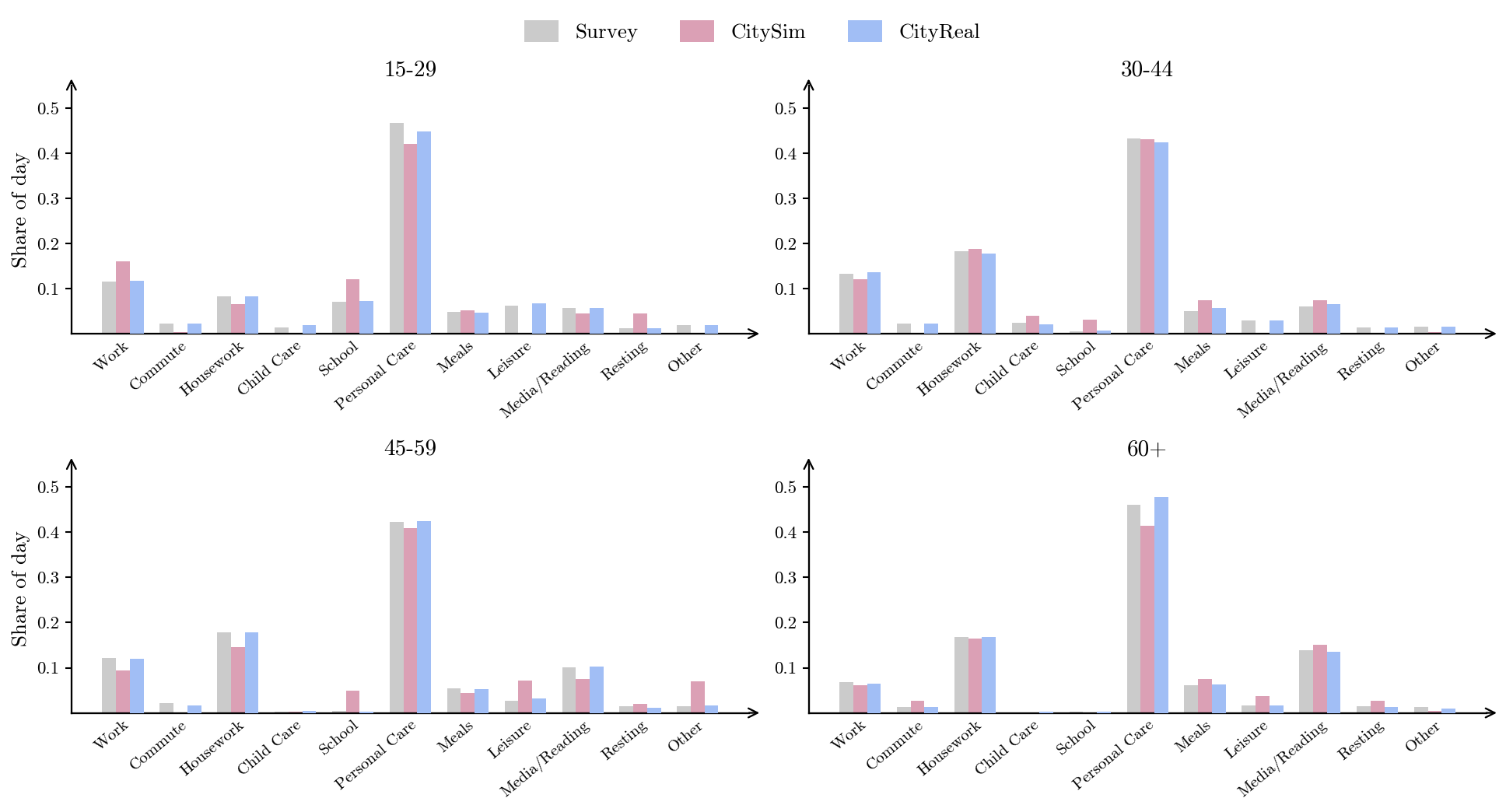}
    \caption{
    Time-use distributions by age group.
    }
    \label{fig:timeuse_compact}
\end{figure*}

This experiment examines whether \textsc{CityReal} improves alignment within demographic groups, rather than only matching aggregate population statistics. Agents are grouped by age and compare their simulated time-use distributions with the 2021 Japanese national time-use survey~\cite{e-stat2021timeuse}. For each bin, we compute the Jensen--Shannon divergence between the simulated and survey activity distributions, where lower values indicate closer alignment. As shown in Figure~\ref{fig:timeuse_compact}, \textsc{CityReal} reduces divergence across all age groups compared with \textsc{CitySim}. The alignment stage preserves age-specific behavioral patterns, rather than improving the population average by collapsing heterogeneous groups into a single distribution.

\subsection{Adapter Convergence and Efficiency}
\label{sec:exp_adapter_convergence}
\begin{figure}[tbp]
    \centering
    \includegraphics[width=1.0\linewidth]{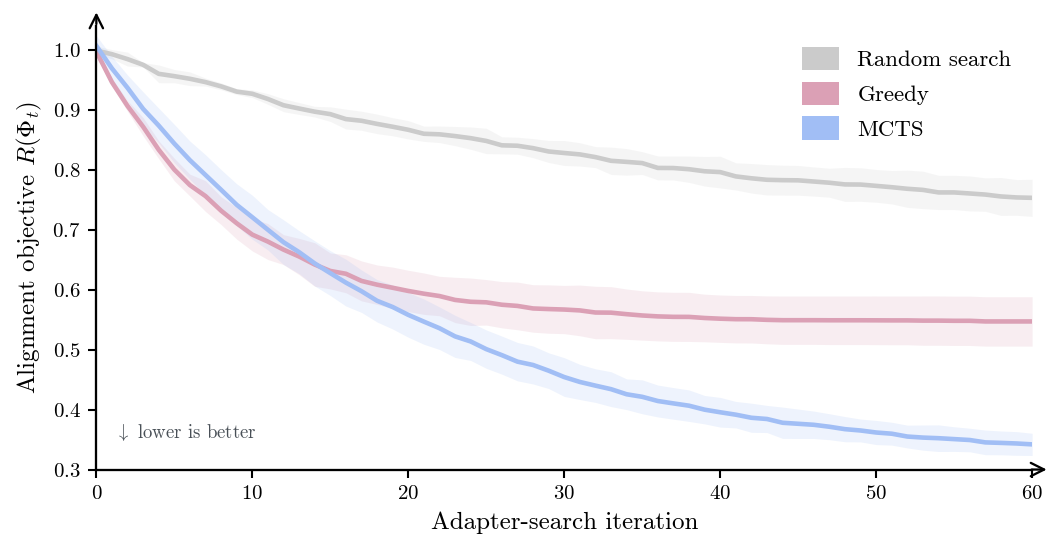}
    \caption{
    Adapter convergence under the same evaluation budget. We plot the best-so-far alignment objective $R(\Phi_t)$. Lower is better.
    }
    \label{fig:adapter_convergence}
\end{figure}

Next, we evaluate whether the adapter search procedure effectively reduces the population-level alignment objective. Starting from empty adapters, we compare MCTS with two simpler alternatives under the same evaluation budget: random search, which samples adapter edits uniformly, and greedy search, which applies the best local edit at each iteration. We report the best-so-far value of $R(\Phi_t)$ across search iterations, where lower values indicate better alignment. MCTS reaches the lowest final objective, as illustrated in Figure~\ref{fig:adapter_convergence}. Greedy search improves quickly at first but then plateaus, suggesting that local edits can correct salient discrepancies but are less effective at coordinating multiple modules and population subgroups. Random search improves more slowly under the same budget. These results support the use of tree search for adapter calibration, as MCTS can explore alternative edit sequences while reusing value estimates from earlier branches.

\subsection{Robustness to Evaluator Bias}
\label{app:evaluator_bias}

\begin{figure}[tbp]
    \centering
    \includegraphics[width=1.0\linewidth]{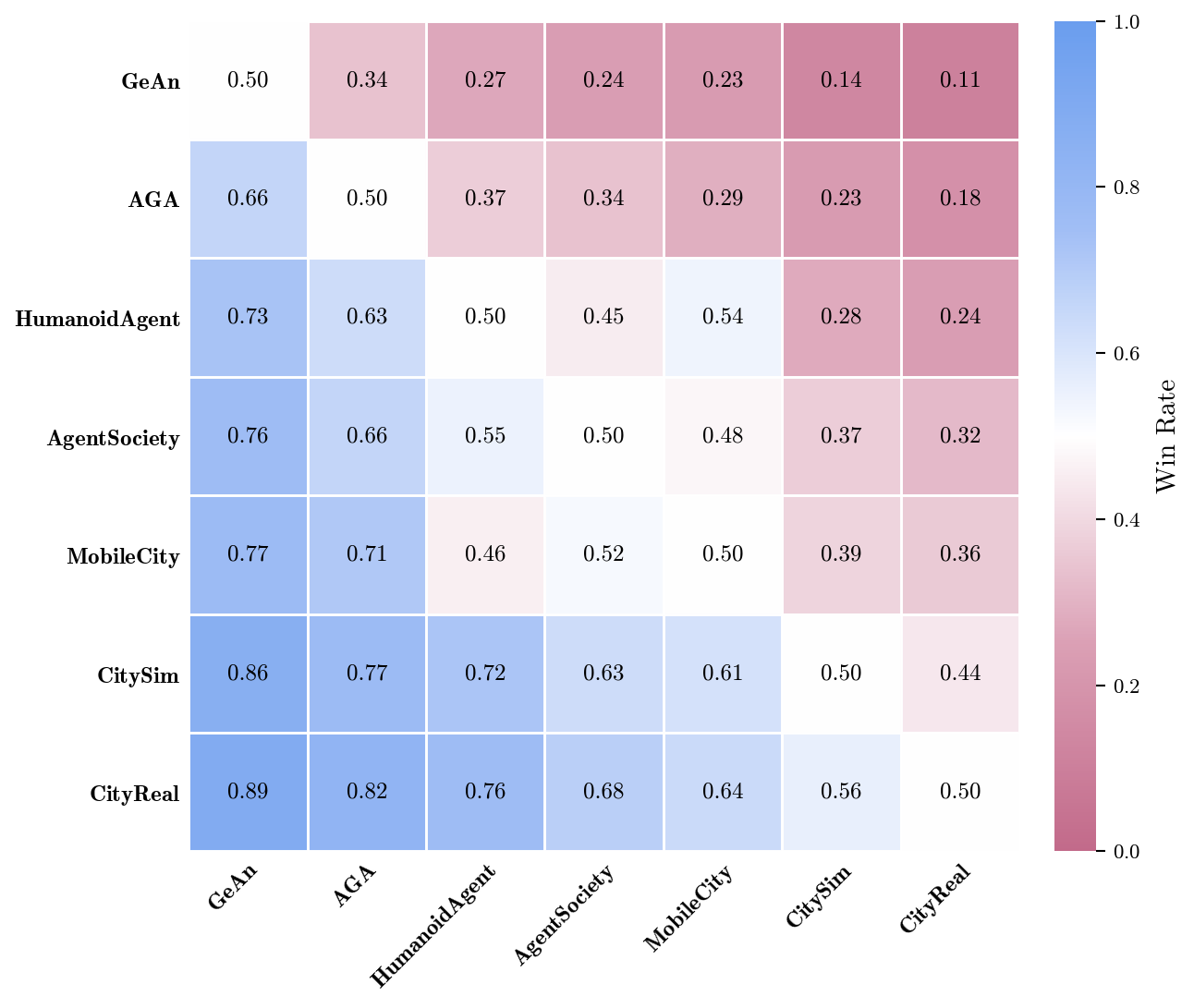}
    \caption{
    Pairwise win-rate matrix using Gemini-3.5 as the evaluator. Each entry indicates the fraction of comparisons in which the row framework is judged more human-like than the column framework.
    }
    \label{fig:winrate_gemini}
\end{figure}

Our main pairwise evaluation uses GPT to judge which simulated behavior appears more human-like. Since \textsc{CityReal} also relies on OpenAI models for agent generation, this setup may raise concerns that the evaluator shares model-family biases with the system being evaluated. To assess this risk, we repeat the same win-rate experiment with Gemini-3.5, an evaluator from a different model family. Figure~\ref{fig:winrate_gemini} depicts that the overall ranking remains stable: \textsc{CityReal} is still preferred over prior frameworks, including \textsc{CitySim}. The margin over \textsc{CitySim} is modest, but consistent, suggesting that the main human-likeness results are not solely driven by evaluator bias from using an OpenAI judge.

\subsection{Urban Policy A/B Testing}
\label{sec:exp_policy_ab}

\begin{figure}[tbp]
    \centering
    \includegraphics[width=1.0\linewidth]{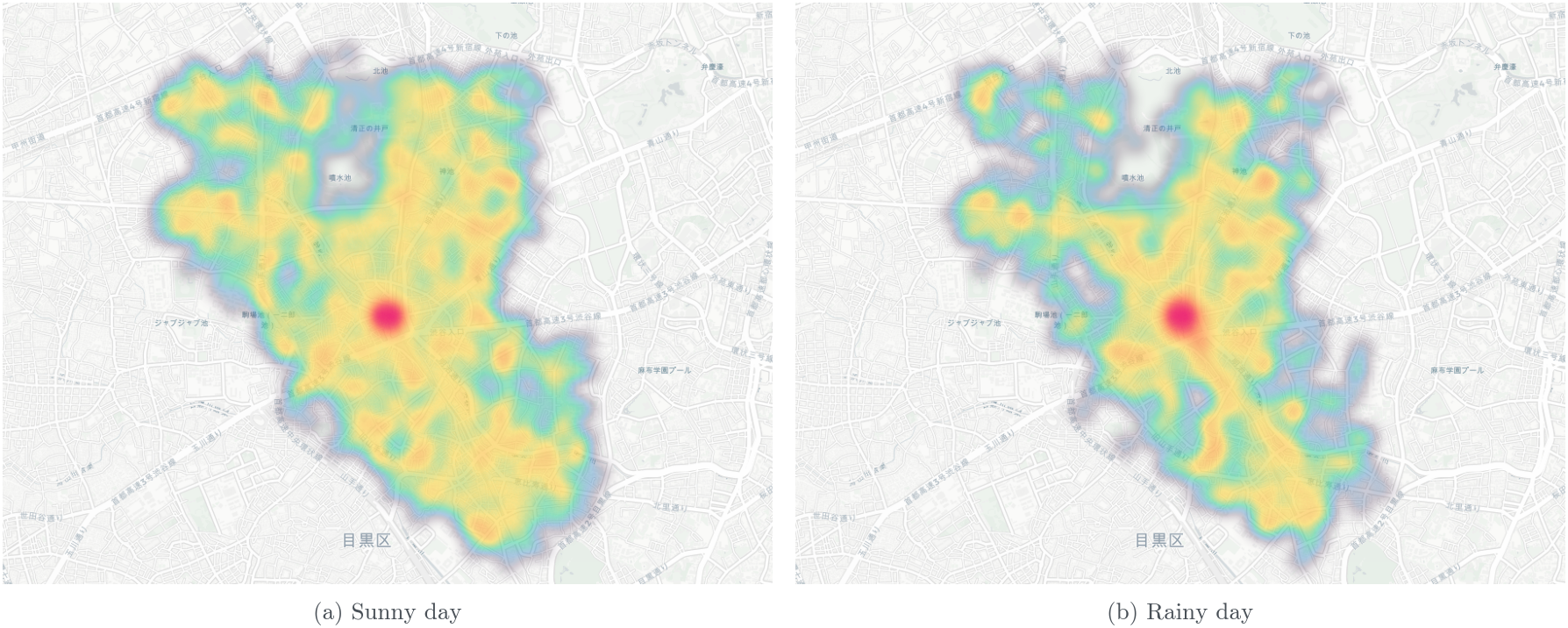}
    \caption{
    Counterfactual weather analysis in Shibuya. Using the same agents and environment, we compare cloudy and rainy conditions. Rain shifts activity from outdoor leisure and walking toward indoor, stay-home, and station-adjacent areas.
    }
    \label{fig:weather_ab}
\end{figure}

We examine whether \textsc{CityReal} supports counterfactual scenario analysis by simulating two Shibuya (Tokyo) scenarios under cloudy and rainy conditions. Because the scenario differs only in weather, changes in the simulated outcomes reflect how agents adapt their plans to environmental conditions. As highlighted in Figure~\ref{fig:weather_ab}, rain shifts agents away from outdoor discretionary activities toward indoor, covered, or stay-home alternatives, while mandatory routines such as work and commuting remain comparatively stable. This scenario illustrates the practical use of \textsc{CityReal} as a counterfactual planning tool. By estimating how weather changes demand, crowding, and movement patterns, the framework can help planners and operators evaluate event plans, station-area congestion, retail foot traffic, and infrastructure stress under alternative urban conditions.

\subsection{Ablation Study}
\label{sec:ablation}

\begin{table*}[tbp]
\centering
\small
\setlength{\tabcolsep}{4.5pt}
\resizebox{1.0\textwidth}{!}{
\begin{tabular}{lcccccccccccccccc c}
\toprule
                 & \multicolumn{15}{c}{Human-likeness (GPT-5, Likert 1--5, $\uparrow$)} & \multicolumn{2}{c}{Population JSD ($\downarrow$)} \\
\cmidrule(lr){2-16} \cmidrule(lr){17-18}
                 & \multicolumn{3}{c}{Activity} & & \multicolumn{3}{c}{Dialogue} & & \multicolumn{3}{c}{Mobility} & & \multicolumn{3}{c}{Event Reaction} & Time-use & Transport \\
\cmidrule(lr){2-4} \cmidrule(lr){6-8} \cmidrule(lr){10-12} \cmidrule(lr){14-16}
\midrule
\rowcolor{blue!10}
CityReal (full) & 4.56 $\pm$ 0.14 & \chart{1273}{97}{cyan} & \texttwemoji{1st_place_medal} &
                & 4.32 $\pm$ 0.05 & \chart{1202}{77}{magenta} & \texttwemoji{1st_place_medal} &
                & 4.46 $\pm$ 0.13 & \chart{1244}{88}{yellow} & \texttwemoji{1st_place_medal} &
                & 4.41 $\pm$ 0.13 & \chart{1229}{84}{green} & \texttwemoji{1st_place_medal} & \textbf{0.0016} & \textbf{0.0021} \\
\midrule
w/o Alignment   & 4.43 $\pm$ 0.17 & \chart{1235}{86}{cyan} & \texttwemoji{2nd_place_medal} &
                & 4.27 $\pm$ 0.06 & \chart{1187}{72}{magenta} & \texttwemoji{2nd_place_medal} &
                & 4.27 $\pm$ 0.16 & \chart{1187}{72}{yellow} & \texttwemoji{2nd_place_medal} &
                & 4.19 $\pm$ 0.15 & \chart{1164}{66}{green} & \texttwemoji{2nd_place_medal} & 0.066 & 0.058 \\
w/o Intention   & 4.08 $\pm$ 0.22 & \chart{1131}{57}{cyan} & &
                & 4.12 $\pm$ 0.07 & \chart{1143}{60}{magenta} & &
                & 3.95 $\pm$ 0.23 & \chart{1092}{46}{yellow} & &
                & 4.00 $\pm$ 0.19 & \chart{1107}{50}{green} & & 0.0058 & 0.0055 \\
w/o Reflection  & 3.95 $\pm$ 0.25 & \chart{1092}{46}{cyan} & &
                & 4.02 $\pm$ 0.07 & \chart{1113}{52}{magenta} & &
                & 3.97 $\pm$ 0.21 & \chart{1098}{47}{yellow} & &
                & 3.88 $\pm$ 0.22 & \chart{1072}{40}{green} & & 0.0044 & 0.0041 \\
w/o Financial   & 4.30 $\pm$ 0.21 & \chart{1196}{75}{cyan} & \texttwemoji{3rd_place_medal} &
                & 4.23 $\pm$ 0.06 & \chart{1175}{69}{magenta} & \texttwemoji{3rd_place_medal} &
                & 4.13 $\pm$ 0.18 & \chart{1146}{61}{yellow} & \texttwemoji{3rd_place_medal} &
                & 4.19 $\pm$ 0.17 & \chart{1164}{66}{green} & \texttwemoji{3rd_place_medal} & 0.0049 & 0.0112 \\
\midrule
w/o Persona     & 3.66 $\pm$ 0.28 & \chart{1006}{22}{cyan} & &
                & 3.55 $\pm$ 0.12 & \chart{974}{20}{magenta} & &
                & 3.72 $\pm$ 0.25 & \chart{1024}{27}{yellow} & &
                & 3.60 $\pm$ 0.24 & \chart{989}{20}{green} & & 0.0091 & 0.0083 \\
w/o Needs       & 3.92 $\pm$ 0.27 & \chart{1084}{43}{cyan} & &
                & 4.08 $\pm$ 0.08 & \chart{1131}{57}{magenta} & &
                & 4.02 $\pm$ 0.19 & \chart{1113}{52}{yellow} & &
                & 3.85 $\pm$ 0.25 & \chart{1063}{38}{green} & & 0.0063 & 0.0049 \\
w/o Belief      & 4.05 $\pm$ 0.22 & \chart{1122}{54}{cyan} & &
                & 4.10 $\pm$ 0.09 & \chart{1137}{58}{magenta} & &
                & 3.98 $\pm$ 0.23 & \chart{1101}{48}{yellow} & &
                & 3.92 $\pm$ 0.22 & \chart{1084}{43}{green} & & 0.0038 & 0.0035 \\
w/o Rec. Plan   & 4.02 $\pm$ 0.24 & \chart{1113}{52}{cyan} & &
                & 4.14 $\pm$ 0.07 & \chart{1149}{62}{magenta} & &
                & 4.06 $\pm$ 0.20 & \chart{1125}{55}{yellow} & &
                & 4.05 $\pm$ 0.21 & \chart{1122}{54}{green} & & 0.0071 & 0.0038 \\
w/o Social      & 4.26 $\pm$ 0.19 & \chart{1184}{72}{cyan} & &
                & 3.98 $\pm$ 0.08 & \chart{1101}{48}{magenta} & &
                & 4.10 $\pm$ 0.20 & \chart{1137}{58}{yellow} & &
                & 4.03 $\pm$ 0.18 & \chart{1116}{53}{green} & & \underline{0.0026} & \underline{0.0029} \\
\bottomrule
\end{tabular}}
\caption{
Ablation study for \textsc{CityReal}. Human-likeness is rated by GPT-5 (Likert, 1--5, mean $\pm$ std; higher is better) for activity, dialogue, mobility, and event reaction; medals denote the top-3 configurations per domain. The last two columns report population-level alignment as the Jensen--Shannon divergence between simulated and survey distributions for time use (calendar) and transport distribution (vehicle), where lower is better.}
\label{tab:ablation_cityreal}
\end{table*}

This section examines the contribution of each component of \textsc{CityReal}. Starting from the full model, we remove one component at a time: population-level alignment, financial pressure, social interaction, experience-driven reflection, intention formation, recursive planning, beliefs, needs, and persona. We evaluate both \emph{individual} realism, measured by human-likeness across four domains, and \emph{population-level} alignment, measured by the Jensen--Shannon divergence between simulated and survey distributions for time use and transport-mode share. Table~\ref{tab:ablation_cityreal} reports the results. The two evaluation axes capture different aspects of realism. Removing alignment mainly affects population-level metrics, substantially increasing the divergence for both time use and transport-mode share. It also reduces human likeness across all domains, suggesting that the learned adapters improve not only aggregate calibration but also individual behavioral realism by preventing agents from reverting to generic LLM priors. Other ablations keep the alignment stage active and therefore remain close to the calibrated population targets, while primarily affecting individual coherence. Among the CityReal components, removing financial pressure degrades mobility and transport-mode alignment, as agents lose an important socioeconomic signal for destination and mode choice. Removing intention mainly affects mobility and event reaction, as without persistent motives, flexible activities become disconnected one-step decisions, weakening trajectory continuity. Disabling reflection produces large drops in activity and event reaction, since agents can no longer accumulate habits and preferences from experience. Removing social interaction mainly affects dialogue and event reaction, while activity is less affected. Besides, our alignment technique also reduces human likeness across all domains, suggesting that the learned adapters improve not only aggregate calibration but also individual behavioral realism by preventing agents from reverting to generic LLM priors. In contrast, all other ablations retain the alignment stage and therefore remain relatively close to the target population distributions.

\subsection{Effect of Search Subset Size}
\label{sec:exp_subset_size}

\begin{table}[tbp]
\centering
\small
\begin{tabular}{lccc}
\toprule
Subset & Cov. & Time-use JSD & Transport JSD \\
\midrule
100            & 3\%   & 0.0072 $\pm$ 0.0019 & 0.0103 $\pm$ 0.0026 \\
250            & 8\%   & 0.0034 $\pm$ 0.0011 & 0.0050 $\pm$ 0.0015 \\
\rowcolor{blue!10}
500            & 17\%  & 0.0016 $\pm$ 0.0004 & 0.0021 $\pm$ 0.0006 \\
1000           & 33\%  & 0.0012 $\pm$ 0.0003 & 0.0015 $\pm$ 0.0004 \\
3000 (full)    & 100\% & 0.0010 $\pm$ 0.0002 & 0.0012 $\pm$ 0.0003 \\
\bottomrule
\end{tabular}
\caption{
Effect of alignment search subset size on population-level alignment for a $3{,}000$-agent population. ``Cov.'' denotes the subset coverage. We report Jensen--Shannon divergence for time use and transport-mode share. Lower is better.
}
\label{tab:subset_size}
\end{table}
The alignment search is performed on a subset of agents, and the learned adapters are transferred to the remaining population. Increasing the subset size consistently improves population-level alignment, as reported in Table~\ref{tab:subset_size}. Small subsets underrepresent minority persona groups, leading to weaker calibration, especially for transport-mode distribution. Gains diminish once the main persona types are sufficiently covered, suggesting that full-population search provides only limited additional benefit. We therefore use $500$ agents in our experiments, which achieves most of the alignment improvement while limiting search cost.

\section{Cost Analysis}
\label{app:efficiency}
We report the LLM cost and scalability of \textsc{CityReal} and compare it with CitySim, the closest baseline. Table~\ref{tab:cost} presents the daily token usage and estimated cost per 1,000 agents, using GPT-5.4-mini pricing. Because \textsc{CityReal} includes population-level alignment, we also report the one-time cost of adapter calibration. In our default setting, calibration is performed on a subset of agents with short simulation rollouts, costing approximately \$248. This cost is incurred once for a target population and can be reused across agents, simulation days, and scenario analyses. The recurring daily cost of \textsc{CityReal} remains comparable to scalable LLM-agent baselines and substantially lower than prompt-heavy simulators such as GeAn. Costs scale approximately linearly with the agent population, and the daily cost for 1,000 agents remains much lower than involving real humans in comparable social studies.

\begin{table}[tbp]
\centering
\small
\resizebox{1.0\columnwidth}{!}{
\begin{tabular}{lcc}
\toprule
Method & Tokens/day (M) & Cost/day (USD) \\
\midrule
CitySim        & 16.71 & \$5.51 \\
AgentSociety   & 14.98 & \$4.82 \\
MobileCity     & 14.41 & \$3.46 \\
AGA            & 22.05 & \$7.01 \\
GeAn           & 69.84 & \$23.60 \\
\rowcolor{blue!10}
CityReal (simulation only) & 19.58 & \$6.76 \\
\bottomrule
\end{tabular}}
\caption{
Daily LLM token usage and estimated cost per 1,000 agents using GPT-5.4-mini.}
\label{tab:cost}
\end{table}

\section{Comparison with Prior Work}
\label{app:comparison}
This section provides a systematic comparison of \textsc{CityReal} against recent agent-based urban simulation systems. Table~\ref{tab:framework_comparison} summarizes agent modeling dimensions, including alignment, adaptive memory, intention modeling, financial state, social interaction, and scalability. Table~\ref{tab:mobility_comparison} focuses on spatial and mobility reasoning.

\begin{table*}[tbp]
\centering
\small
\resizebox{1.0\textwidth}{!}{
\begin{tabular}{lcccccccl}
\toprule
Method 
& Pop. Align. 
& Rich Persona 
& Needs 
& Economy 
& Adaptive Memory 
& Persistent Intent. 
& Social 
& Scale \\
\midrule

\rowcolor{blue!10}
\textsc{CityReal} 
& \checkmark 
& \checkmark 
& \checkmark 
& \checkmark 
& \checkmark 
& \checkmark 
& F2F + online 
& $>$10k \\

CitySim 
& \xmark 
& \checkmark 
& \checkmark 
& \xmark 
& partial 
& \xmark 
& F2F + remote 
& $>$10k \\

AgentSociety 
& \xmark 
& partial 
& \checkmark 
& \xmark 
& partial 
& \xmark 
& F2F 
& $>$10k \\

MobileCity 
& \xmark 
& partial 
& \xmark 
& \xmark 
& partial 
& \xmark 
& F2F 
& 4k \\

GeAn 
& \xmark 
& partial 
& \xmark 
& \xmark 
& partial 
& \xmark 
& F2F 
& 25 \\

AGA 
& \xmark 
& partial 
& \xmark 
& \xmark 
& partial 
& \xmark 
& F2F 
& 100 \\

HumanoidAgent 
& \xmark 
& partial 
& \xmark 
& \xmark 
& \xmark 
& \xmark 
& F2F 
& 100 \\

\bottomrule
\end{tabular}}
\caption{
Comparison of agent-level modeling capabilities. \textsc{CityReal} is the only framework that jointly supports population-level alignment, persistent intentions, financial state modeling, and adaptive memory at large scale.
}
\label{tab:framework_comparison}
\end{table*}

\begin{table*}[tbp]
\centering
\small
\resizebox{1.0\textwidth}{!}{
\begin{tabular}{lcccccc}
\toprule
Method 
& Intent.-Aware Area 
& POI Beliefs 
& Belief Updates 
& Belief-Gravity 
& Transport Choice 
& Cost-Aware Choice \\
\midrule

\rowcolor{blue!10}
\textsc{CityReal} 
& \checkmark 
& \checkmark 
& \checkmark 
& \checkmark 
& \checkmark 
& \checkmark \\

CitySim 
& \xmark 
& \checkmark 
& \checkmark 
& \checkmark 
& \checkmark 
& \xmark \\

AgentSociety 
& \xmark 
& \xmark 
& \xmark 
& \xmark 
& \xmark 
& \xmark \\

MobileCity 
& \xmark 
& \xmark 
& \xmark 
& \xmark 
& rule-based 
& \xmark \\

GeAn 
& \xmark 
& \xmark 
& \xmark 
& \xmark 
& \xmark 
& \xmark \\

AGA 
& \xmark 
& \xmark 
& \xmark 
& \xmark 
& \xmark 
& \xmark \\

HumanoidAgent 
& \xmark 
& \xmark 
& \xmark 
& \xmark 
& \xmark 
& \xmark \\

\bottomrule
\end{tabular}}
\caption{
Comparison of spatial and mobility reasoning. \textsc{CityReal} extends prior mobility models with intention-aware area selection, belief-aware gravity scoring, transport choice, and cost-aware decision making.
}
\label{tab:mobility_comparison}
\end{table*}

\section{LLM Evaluator Prompt}
To assess whether interaction traces resemble those of real users or are indicative of AI-generated behavior, we employ an LLM-based evaluator. This judge is prompted as follows:
\begin{tcolorbox}[colframe=customblues3, colback=white, title=LLM Evaluator Prompt, breakable]
Please evaluate the following urban behavior trace of a simulated agent and determine whether it resembles behavior generated by an AI agent or by a real human:\\
\textcolor{customorange}{\{interaction logs\}}\\

Please rate on a scale of 1 to 5, with 1 being most like an AI and 5 being most like a human. 
\end{tcolorbox}

\section{Simulation Interface}

\begin{figure}[tbp]
    \centering
    \includegraphics[width=1.0\linewidth]{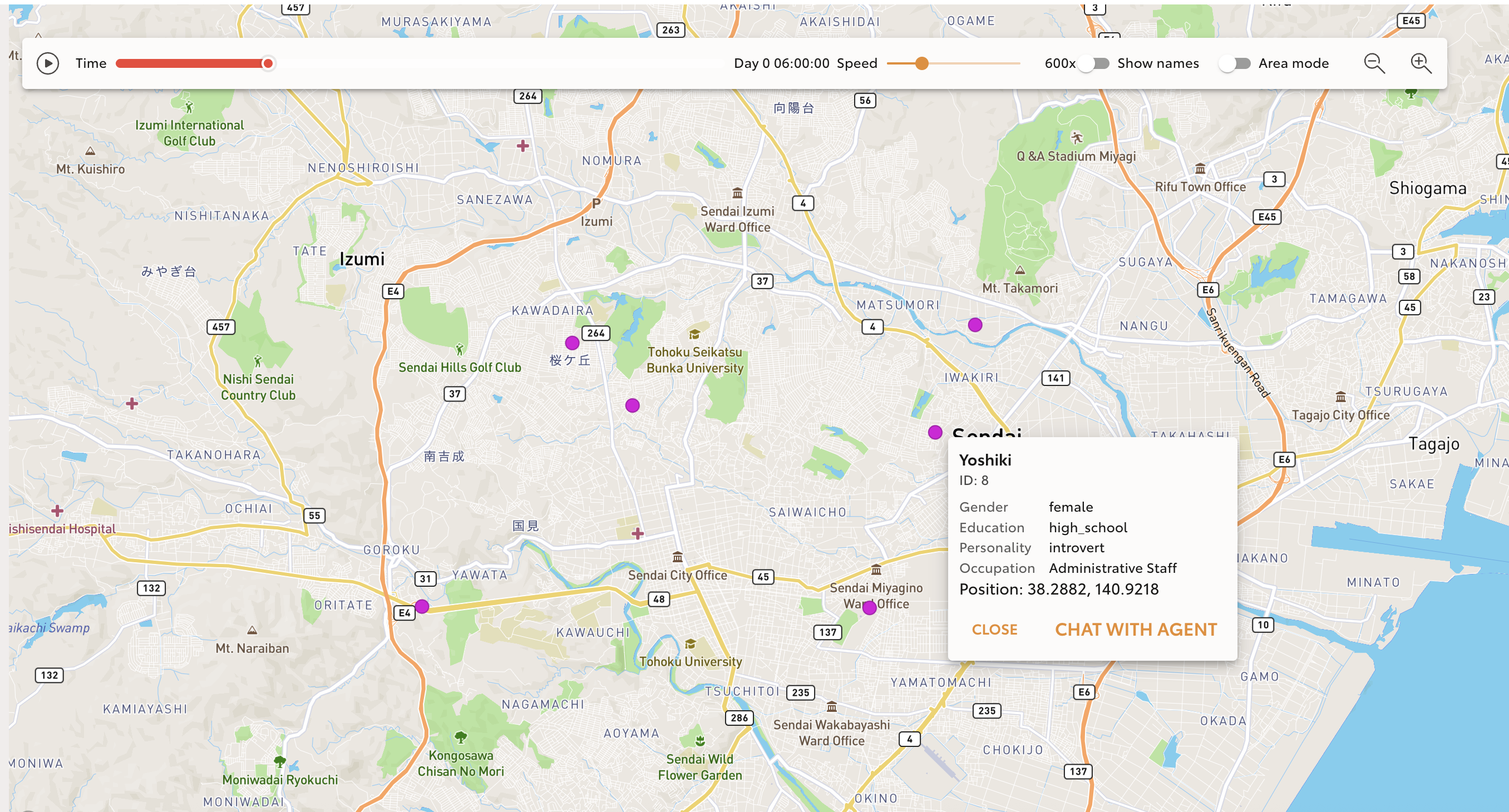}
    \caption{Simulation interface for visualizing agent behavior and city dynamics.}
    \label{fig:interface}
\end{figure}

To support qualitative analysis and scenario exploration, we develop an interactive interface for visualizing agent behavior and city-level dynamics. Figure~\ref{fig:interface} illustrates the interface, which visualizes agent trajectories, activities, intentions, schedules, mobility patterns, and contextual information in real time. It also allows users to inspect individual agents, compare behaviors across populations, and analyze how routines evolve under different environmental conditions or interventions. This interface is designed to support researchers in understanding not only aggregate simulation outcomes, but also the underlying behavioral processes that generate them.

\end{document}